\documentclass[fleqn,usenatbib]{mnras}

\usepackage{newtxtext,newtxmath}

\usepackage[T1]{fontenc}
\usepackage{ae,aecompl}

\usepackage{graphicx}
\usepackage{rotating}
\usepackage{amsmath}
\usepackage{longtable}
\usepackage{float}
\usepackage{soul}
\usepackage{pdflscape}
\usepackage{longtable}

\usepackage{xcolor}

\usepackage{lineno}

\title[A radio-optical study of Abell\,1300 and MACS\,J1931.8--2634]{A MeerKAT view on galaxy clusters: a radio-optical study of Abell\,1300 and MACS\,J1931.8--2634}

\author[B. Terni de Gregory et al.]{B.~Terni de Gregory$^{1,2}$\thanks{E-mail: b.gregory@ira.inaf.it},
B.~Hugo$^{3,4}$,
T.~Venturi$^{1}$,
G.~Bernardi$^{1,3,4}$,
D.~Dallacasa$^{1,2}$,
M.~Nonino$^{5}$,
\newauthor
S.~Makhatini$^{3}$,
V.~Parekh$^{3}$,
O.M.~Smirnov$^{3,4}$,
S.~Giacintucci$^6$ and
R.~Kale$^7$
\\
\\
$^{1}$INAF-Istituto di Radio Astronomia, via Gobetti 101, 40129 Bologna, Italy\\
$^{2}$Dipartimento di Fisica e Astronomia, Universit\'{a} di Bologna, Via Gobetti 93/2, 40129 Bologna, Italy\\
$^{3}$South African Radio Astronomy Observatory, Black River Park, 2 Fir Street, Observatory, Cape Town, 7925, South Africa\\
$^{4}$Department of Physics and Electronics, Rhodes University, PO Box 94, Grahamstown, 6140, South Africa\\
$^{5}$INAF-Osservatorio Astronomico di Trieste, Via G.B. Tiepolo 11, I-34143 Trieste, Italy\\
$^6$Naval Research Laboratory, 4555 Overlook Avenue SW, Code 7213, Washington, DC 20375, USA\\
$^7$National Centre for Radio Astrophysics, Tata Institute of Fundamental Research, S. P. Pune University Campus, Ganeshkhind, Pune 411007, India
}

\date{Accepted XXX. Received YYY; in original form ZZZ}

\pubyear{2020}

\begin{document}
\label{firstpage}
\pagerange{\pageref{firstpage}--\pageref{lastpage}}
\maketitle

\begin{abstract}
In this paper we present results from a radio-optical study of the galaxy populations of the galaxy clusters Abell~1300 and MACS\,J1931.8$-$2634, a merger and a relaxed system respectively both located at $z \sim 0.3$,
aimed at finding evidence of merger-induced radio emission. Radio observations are taken at 1.28~GHz with the MeerKAT interferometer during its early-stage commissioning phase, and combined with archive optical data.
We generated catalogues containing 107 and 162 radio sources in the A\,1300 and MACS\,J1931.8--2634 cluster fields respectively, above a 0.2~mJy threshold and within a 30~arcmin radius from the cluster centre (corresponding to 8.1 and 8.8~Mpc respectively). By cross-correlating the radio and optical catalogues, and including spectroscopic information, 9 and 6 sources  were found to be cluster members and used to construct the radio luminosity functions respectively for both clusters. The comparison of the radio source catalogues
between the two cluster fields leads to a marginal difference, with a $2\sigma$ statistical significance.
We derived the radio luminosity function at 1.28 GHz in both clusters, in the power range $22.81 < \rm {log~P_{1.28~GHz}~(W/Hz)} < 25.95$, and obtained that in A\,1300 the radio luminosity function averaged over the full radio power interval is only $3.3 \pm 1.9$ times higher than the MACS\,J1931.8--2634 one, suggesting no statistical difference in their probability to host nuclear radio emission.
We conclude that, at least for the two clusters studied here, the role of cluster mergers in affecting the statistical properties of the radio galaxy population is negligible. 
\end{abstract}




\section{Introduction}

Cluster mergers are the most energetic events in the Universe. They are the natural way of forming rich clusters of galaxies within cold dark matter scenarios, which imply a bottom-up hierarchy of structure formation \citep{Press74,Sarazin02}. The merging process generates important perturbations in the intra-cluster medium such as shocks, bulk flows and turbulence in the hot gas, which considerably affect the properties of the non-thermal components of galaxy clusters. As a matter of fact it is nowadays well established that the spectacular cluster-scale radio emission in the form of halos and relics originates by merging processes \citep[i.e.,][and references therein]{Bruggen12,Brunetti14,Botteon16}. 

While the morphology of radio galaxies is strongly affected by the environment, as witnessed by the presence of wide--angle tails (WAT) and narrow--angle tails (NAT) in the majority of galaxy clusters and groups \citep[see][for a review on the topic]{Feretti02},
the role of cluster formation processes on the statistical radio properties of the galaxy population is still unclear. 
The most powerful tool to investigate the influence of the environment on the radio properties of galaxies is the Radio Luminosity Function (RLF), which gives the probability for a galaxy to develop a radio source above a given radio power threshold. The comparison of RLFs for galaxies in different environments and at different redshifts provides clues on the role of the environment and of the cosmological evolution.
To date, results of these studies are however inconclusive. 
The few clusters studied in detail so far in the local Universe provide conflicting results both for the radio nuclear activity and for the starburst population of galaxies \citep{Owen99,Dwarakanath99,Venturi00,Venturi01,Miller03,Giacintucci04}, suggesting that several parameters might play a relevant role. 

The cosmological evolution of the RLF is a matter of debate, too.
Studies of clusters at intermediate redshift ($0.3 < z < 0.8$) provide different results for different samples
\citep{Stocke99,Branchesi06,Gralla11} when compared to the local radio luminosity function of cluster galaxies \citep{Ledlow96}.

The only solid results are the dependence of the RLF on the optical magnitude of the host galaxy, at least up to $z = 0.3$ \citep{Mauch07}, and the remarkably different behaviour of the  Brightest Cluster Galaxies (BCG) in relaxed and merging clusters in the GMRT Radio Halo cluster sample \citep{Kale15} and for the HIFLUG cluster sample at lower redshift \citep{Mittal09}. These show that the fraction of brightest cluster galaxies with radio emission is much higher in relaxed clusters at least up to $z = 0.4$, and increases with increasing cool-core strength.

During the MeerKAT-16 commissioning stage, we observed the two galaxy clusters Abell 1300 (A\,1300) and MACS\,J1931.8--2634. While the main goal of the observations was to test the telescope performance and new calibration and imaging procedures, the galaxy clusters were selected with the scientific aim to address the role of cluster mergers in shaping the radio properties of cluster radio galaxies.
The two clusters are at similar redshifts ($z \sim 0.3$) and have similar mass, but they differ in their dynamical properties: A\,1300 is considered a post--merger, while MACS\,J1931.8--2634 is classified as a relaxed system.
The two clusters were observed at 1.28~GHz with the MeerKAT array, and the radio observations were complemented with optical {\it Subaru}-SuprimeCam images for a comparative study of the radio galaxy population in two different environments, which we carried out comparing their radio source counts and their RLFs.

In this paper we present the results of our study. The layout is as follows: Section~\ref{sec:obz} describes the observations and data reduction; the optical data are described in Section~\ref{sec:optical}; radio catalogues, source counts, optical IDs and RLFs are shown in Section~\ref{sec:results} and conclusions are offered in Section~\ref{sec:discussion}. 
Throughout the paper we assume $H_0 = 70$~km~s$^{-1}$~Mpc$^{-1}$, $\Omega_m = 0.3$ and $\Omega_\Lambda = 0.7$, which give a 4.5~kpc~arcsec$^{-1}$ and 4.9~kpc~arcsec$^{-1}$ scale for A\,1300 and MACS\,J1931.8--2634 respectively.

\subsection{A~1300}

The galaxy cluster A\,1300 is reported as a $z = 0.31$, richness class 1 object by \citet{Abell89}. It was first surveyed in the X-ray band  during the Rosat All Sky Survey \citep{Pierre94}, where it is identified with RXJ1131.9-1955. It has an X-ray luminosity of $L_x \sim 1.7 \times 10^{45}$~erg~s$^{-1}$ in the $0.1-2.4$~keV band  and a mass M$\sim 1.3 \times 10^{15}$~M$_{\odot}$ \citep{Lemonon97}. Its virial radius is 1.53~Mpc, corresponding to 5.7$^{\prime}$ \citep{Ziparo12}.

Optical \citep{Pierre97} and X-ray \citep{Lemonon97} observations suggest that A\,1300 is in a post-merging phase. It is estimated that a major merger occurred about 3 Gyr ago, with further accretion taking place along the cosmic web filaments, leading to an increase of the cluster mass up to 60\% in the next $\sim$Gyr \citep{Ziparo12}. \cite{Ziparo12} quote 987 km~s$^{-1}$ as rest-frame velocity dispersion, which they also use to estimate a dynamical mass of $M_{200} \approx 1.1 \times 10^{15} M_{\odot}$. 
The cluster hosts a giant radio halo and a relic located in the south--western periphery of the cluster and a number of extended radio galaxies in the central regions \citep{Reid99,Venturi13}.
A\,1300 is among the targets of the Merging Cluster Collaboration (MCC)\footnote{http://www.mergingclustercollaboration.org/},  and of the Galaxy Cluster at VirCam Survey (GCAV)\footnote{https://www.eso.org/sci/observing/PublicSurveys/docs/GCAV}, an Infra-red, 560 hrs, ESO Public Survey (PI Nonino M.) in the Y, J and Ks bands, whose aim is to explore galaxy evolution over a large variety of environments.

\subsection{MACS~J1931.8--2634}

This massive ($M_{200} = 1.74\times 10^{15}M_{\odot}$, \cite{Umetsu14}), $z = 0.35$, cool core galaxy cluster \citep{Ebeling10} is part of the Massive Cluster Survey sample \citep[MACS, see][and references therein]{Ebeling07} and has an X-ray luminosity of $L_x \sim 2.2 \times 10^{45}$~erg~s$^{-1}$ in the $0.1-2.4$~keV. It hosts one of the X-ray most luminous cool cores discovered so far, with an equivalent mass cooling rate within the central 50~$h^{-1}_{70}$~kpc of $\sim 700$~M$_{\odot}$~yr$^{-1}$ \citep{Ehlert11}. Its virial radius is 1.8~Mpc, corresponding to 6.1~arcmin \citep{Santos16}. 
\\
\cite{Santos16} report a weak-lensing mass of $M_{200} \approx 0.99 \times 10^{15}$~M$_{\odot}$ \citep[see also][]{Merten15}. 
For comparison with A\,1300, from the mass we estimate a velocity dispersion following \citet{Carlberg97} \citep[using the same cosmology as][]{Ziparo12}, resulting in  $\sigma \approx 900$ km~s$^{-1}$, i.e. similar to A\,1300.  The two clusters are thus quite similar in mass, meaning, via the mass-richness scaling relation \citep{Andreon10,Melchior17}, that they have similar richness.

The BCG is surrounded by X-ray cavities created by an outburst from the central source \citep{Allen04} and is undergoing intense star formation \citep{Fogarty15,Donahue15,Santos16}. Moreover,
it  is embedded in extended radio emission whose nature, i.e. radio lobes or mini--halo, is still a matter of debate \citep{Giacintucci04,Giacintucci17}.
The cluster hosts a Narrow-Angle Tailed (NAT) radio galaxy, located at $\sim 200$~kpc linear projected distance south of the BCG. MACS\,J1931.8-2634 is part of the Cluster Lensing and Supernovae Survey with Hubble sample  \citep[CLASH,][]{Postman12}.

\section{MeerKAT observations and data reduction}\label{sec:obz}

The MeerKAT radio telescope\footnote{https://www.sarao.ac.za} is a precursor for the Square Kilometre Array mid-frequency telescope \citep{Jonas16,Camilo18,Mauch20}. The full array consists of 64 antennas with 13.5--m diameter dishes which operate with an effective bandwidth from 580 to 3500~MHz split into (at present) UHF, L--band and S--band receivers. 
The array configuration consists of a 1--km inner core containing $\approx70\%$ of the dishes and a distribution of antennas reaching a maximum baseline length of 8~km.
The instrumental field of view spans roughly from $0.5^\circ$ to $1^\circ$ at the -20~dB point of the voltage beam, at the low and high ends of the L-band bandwidth \citep[][]{Jonas16,Mauch20}. This wide field of view enables us to image far beyond the virial radii of the two clusters (i.e., $\sim 1.5$~Mpc) for our statistical analysis.

\begin{figure}
\centering
\includegraphics[scale=0.4]{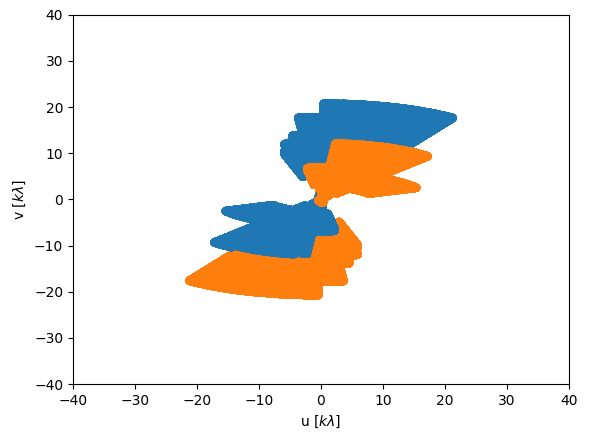}
\includegraphics[scale=0.4]{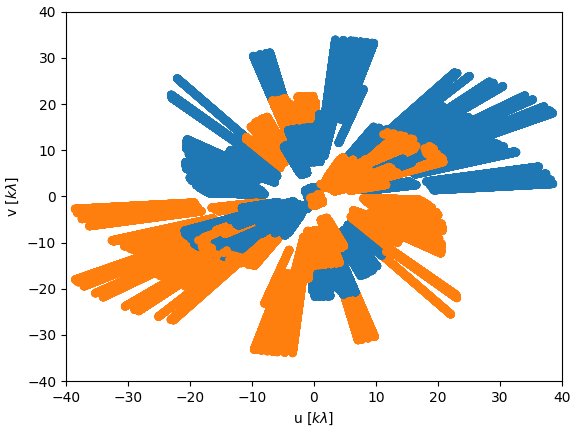}
\caption[The uv coverage]{A\,1300 (top panel) and MACS\,J1931.8--2634 (bottom panel) $uv$--coverages over the entire bandwidth. Only one out of every ten points in time and one out every 40 points in frequency are plotted. The two colours show the symmetric $uv$--points obtained from the conjugate visibilities. Note that the different $uv$-coverage for the two clusters reflects into a different synthesized beam (Table~\ref{tab:Details_observations}).}
\label{fig:uv_coverage}
\end{figure}

The observations presented in this paper were performed during the Array Release 1.5 construction phase, i.e. when only 16 antennas were available. Details of the observations are reported in Table~\ref{tab:Details_observations}.
The data were taken at L--band, with a central frequency of 1.283~GHz ($\lambda = 0.23$~m) and a total bandwidth of 852~MHz. The signal correlation was carried out using 4096~channels in total, each 208~kHz wide, with a 2~s integration time. 
Even though the observations of the two clusters were close in time, the array configuration differed, and this is reflected into the $uv$--coverage, shown in Fig.~\ref{fig:uv_coverage} for both sources. The combination of the different $uv$-coverage and more severe RFI excision led to a worse sampling of the $uv$ plane and lower angular resolution for A\,1300
(see also Table~\ref{tab:Details_observations}).
\begin{table*}
\caption[Details on the radio observations]{Details of the radio  observations. From left to right: (1) observing date, (2) target name, (3) and (4) J2000 coordinates, (5) central observing frequency ($\nu_c$), (6) bandwidth, (7) integration time, (8) full width half maximum and (9) position angle of the restoring beam, (10) image noise.}
\label{tab:Details_observations}
\centering
\begin{footnotesize}
\begin{center}
\begin{tabular}{c c c c c c c c c c}
\hline
\\
(1) & (2) & (3) & (4) & (5) & (6) & (7) & (8) & (9) & (10) \\
Obs. date& name& RA$_{J2000}$& DEC$_{J2000}$&  $\nu_c$&  BW&   int. time&   FWHM&  PA& rms \\
& & h m s & $^\circ$ ' "& MHz& MHz& hr& arcsec~$\times$~arcsec & $\deg$& mJy~beam$^{-1}$ \\
\hline
\\
April 2018& A\,1300&  11:31:54.4& -19:55:42& 1283& 856& 2.4& 12 x 5& 113& 0.04 \\
\\
May 2018& MACS\,J1931.8--2634&  19:31:49.6& -26:34:34& 1283& 856& 2.0& 5 x 3& 136& 0.04 \\
 \hline
 \\
\end{tabular}
\end{center}
\end{footnotesize}
\end{table*}

The data reduction was carried out using the NRAO \texttt{Common Astronomy Software Application (CASA)} package \citep{mcmullin2007casa} for the a-priori calibration and a combination of \texttt{CubiCal} \citep{Kenyon18} and \texttt{WSClean} \citep{Offringa14} for the self-calibration. Moreover, we used \texttt{DDFacet} \citep{Tasse18} to obtain the primary beam corrected images. The packages were pipelined together by means of the containerized \texttt{Stimela} scripting platform \citep{makhathini_phdthesis}. To ensure self consistency between the two fields, we adopted the same strategy for the bandpass calibration, imaging and self-calibration.
The complex bandpass was derived from the observation of PKS~B1934--638 \citep{reynolds1994revised}.
RFI mitigation was accomplished using a combination of a static mask of known L-band interferences including band rolloffs, and residual autoflagging using \texttt{AOFLAGGER} \citep{Offringa12,Offringa13}. Multiple rounds of flagging and calibration were performed to remove residual narrow-band RFI. 

The complex bandpass was applied to the target, and the data were averaged by a factor of two in frequency (obtaining a 416~kHz channel width). The number of channels on which we performed the direction independent self-calibration became 2048.
The final effective bandwidth after RFI removal (which excludes the band edges) is about 750~MHz.

Self-calibration was then performed with a combination of non-parametric time-variable gains using CubiCal and sky models synthesized using the wide-field, wide-band WSClean imager.
The initial model images were deconvolved down to a 0.2~mJy threshold ($5 \sigma$). 
The self-calibration was performed by four iterations with phase-only solutions. The solution interval was reduced inspecting the solutions at each step. Each imaging step in the self-calibration cycle was carried out with  
Briggs weighting scheme -2 \citep{Briggs95} to achieve the maximum resolution needed to fulfill our scientific goals.
Deconvolution was performed using a combination of automatic masking, cycle-variant local RMS thresholding and manual masking to limit artifacts around bright sources. The final images were obtained with a final deconvolution without the manual mask. The images shown in this paper are not primary-beam corrected.
Primary beam corrected images were generated using the DDFacet package \citep{Tasse18} and used for the analysis reported in Section~\ref{sec:results}.
We imaged out to an extent of $\sim 1^\circ$ in diameter using $19 \times 19$ facets for which the rms noise is within 80\% of the central noise level at the top of the bandwidth.
The antenna response used in this patch-wise correction is smoothed from holographic measurements carried out at L--band \citep{asad2019primary}.

\section{Optical data}\label{sec:optical}

We used \textit{Subaru}-SuprimeCam images to identify optical counterparts of the MeerKAT radio sources.
The Suprime stacked images for MACS\,J1931.8--2634 in the B, V, Rc, Ic, z filters were retrieved from the STScI CLASH  data products\footnote{https://archive.stsci.edu/prepds/clash/}. On the other hand, the images for A\,1300 were created from our own data reduction of the raw \textit{Subaru}-SuprimeCam images in the  $g'$, $r'$  bands retrieved from the SMOKA-Subaru Archive\footnote{https://smoka.nao.ac.jp/index.jsp, http://mergingclustercollaboration.org/} 
obtained for the Merging Cluster Collaboration, and from the ESO Archive (WFI images B, V, Rc, Programme 084.A-9001).
We used the same methods which were used to create the B, V, Rc, Ic, z Suprime images for MACS\,J1931.8-2634 \citep[e.g.,][]{Nonino09}. Both data sets were photometrically calibrated using matched point-like sources from PanStarr \citep{Chambers16}, accounting for colour terms. Details of the optical observations are given in  Tables~\ref{tab:MACSJ_optical_obs} and \ref{tab:A1300_optical} for MACS\,J1931.8--2634 and A\,1300 respectively. All magnitudes are in the AB system.
\begin{table}
\centering
\caption[MACSJ1931.8-2634 optical]{MACSJ1931.8--2634 optical observations.
}
\label{tab:MACSJ_optical_obs}
\begin{footnotesize}
\begin{center}
\begin{tabular}{cr cr cr cr cr}
\hline
\\
(1) & (2) & (3) & (4) & (5) \\
Observation date & Filter & exposure & FWHM & depth(*) \\
    & & (s) & (arcsec)  \\ 
\hline
\\
2006 \& 2012& B& 2640& 1.20& 25.98 \\
\hline
\\
2006& V& 1275& 0.88& 25.33 \\
\hline
\\
2006 \& 2012& Rc& 4560& 0.81& 25.18 \\
\hline
\\
2006& Ic& 1800& 0.92& 24.7 \\
\hline
\\
2012& z& 1950& 0.76& 24.42 \\
\hline
\end{tabular}
\end{center}
\end{footnotesize}
{\scriptsize (*) depth: AB magnitude 5-sigma in 2~arcsec-diameter aperture.}
\end{table}

For each cluster, the galaxy catalogues were then generated, and the magnitudes were corrected for galactic extinction according to \cite{Schlafly11}.
The filter sets for the two clusters were not the same, in particular the two $r$ bands throughput are slightly different: using SED fitting of galaxies which are optical counterparts of radio sources 
(see Section~\ref{sec:optical_identifications}) the difference $r'$--Rc ranges from $\approx 0.1$ to $\approx 0.16$ for blue and red galaxies respectively at $z\approx 0.30-0.35$, i.e. at the cluster redshift.
\begin{table}
\centering
\caption[A1300 optical]{A1300 optical observations}\label{tab:A1300_optical}
\begin{footnotesize}
\begin{center}
\begin{tabular}{cr cr cr cr cr}
\hline
\\
(1) & (2) & (3) & (4) & (5) \\
Observation date& Filter& exposure& FWHM & depth* \\
  &  & (s) & (arcsec)& \\ 
\hline
\\
2014& $g'$ & 720& 1.06& 25.69 \\
\hline
\\
2014& $r'$ & 2880& 0.94& 26.17 \\
\hline
\end{tabular}
\end{center}
\end{footnotesize}
{\scriptsize (*) depth: AB magnitude 5-sigma in 2~arcsec-diameter aperture.}\end{table}

\section{Results and Discussion}
\label{sec:results}

\begin{figure*}
    \centering
    \includegraphics[width=0.9\textwidth]{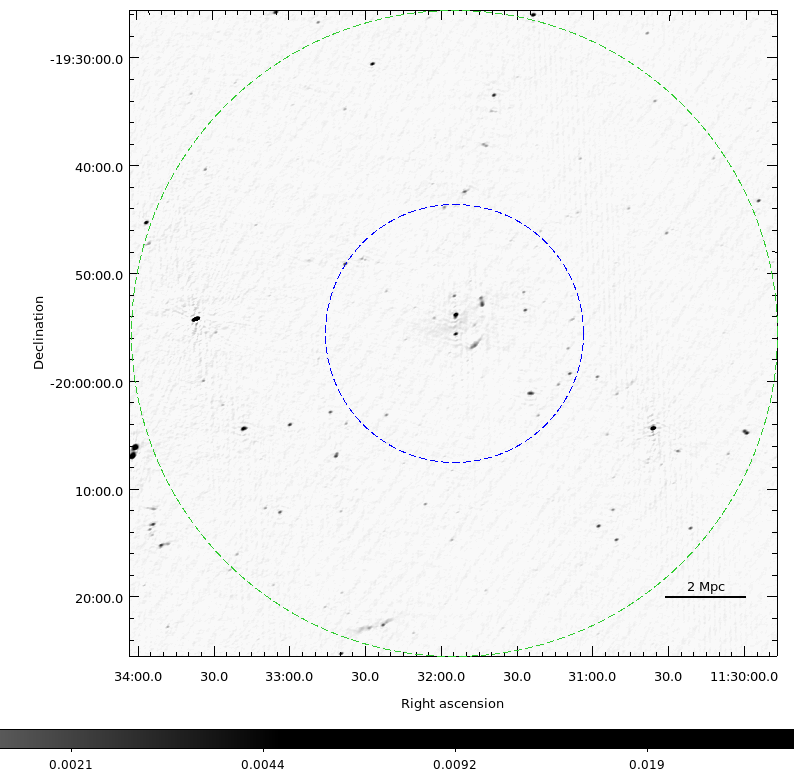}
    \caption[A1300 field]{Gray scale image of the galaxy cluster A\,1300 and its field at 1.28~GHz. The green dashed circle has a $\sim$1~deg diameter and represents the HPBW of the MeerKAT primary beam. This is the area used to extract the source catalogue. The blue dashed circle marks the inner 12~arcmin radius (i.e., 0.125~deg$^2$ area), corresponding to about two virial radii. The angular resolution is 12~arcsec~$\times$~5~arcsec. Units are Jy~beam$^{-1}$.}
    \label{fig:A1300_field}
\end{figure*}

\begin{figure*}
    \centering
    \includegraphics[width=0.9\textwidth]{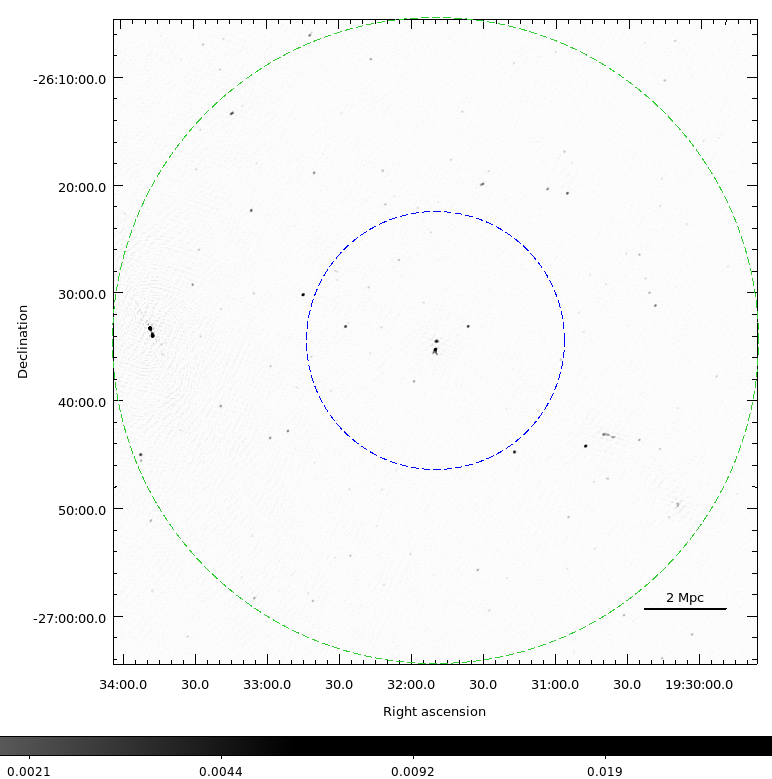}
    \caption[MACSJ1931 field]{Same as Fig.~\ref{fig:A1300_field}, but for the MACS\,J1931.8--2634 galaxy cluster and its field. The angular resolution is 5~arcsec~$\times$~3~arcsec.}
    \label{fig:MACSJ1931_field}
\end{figure*}

The $1^\circ \times 1^\circ$ MeerKAT images are shown in Fig.~\ref{fig:A1300_field} and \ref{fig:MACSJ1931_field}. 
An average rms noise of 40~$\mu$Jy~beam$^{-1}$  is achieved on both targets. 
The image quality is generally good, although affected by some residual stripes likely due to low level residual RFI that were not  entirely removed despite the multiple flagging process. A couple of bright, off--axis sources are affected by calibration errors that would require direction dependent calibration. As they do not impact the goals of the current analysis, we leave this to a future work.

\begin{figure*}
\centering
\includegraphics[scale=0.35]{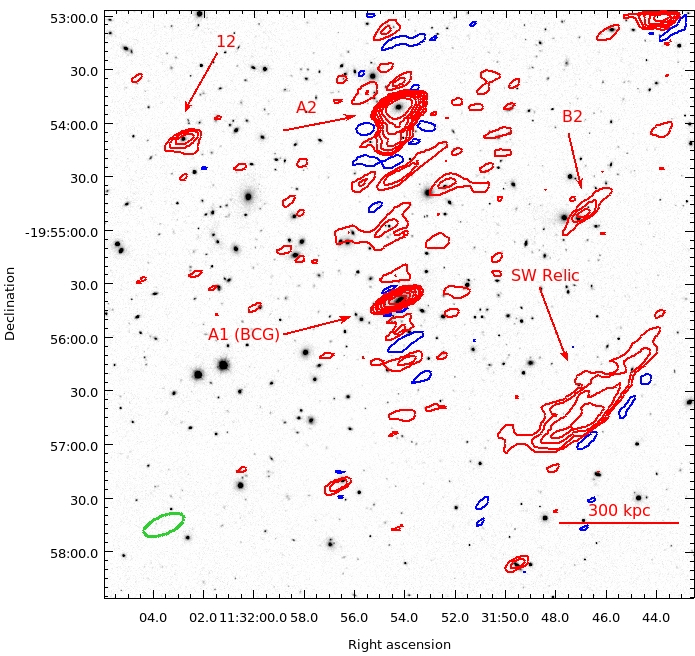}
\includegraphics[scale=0.35]{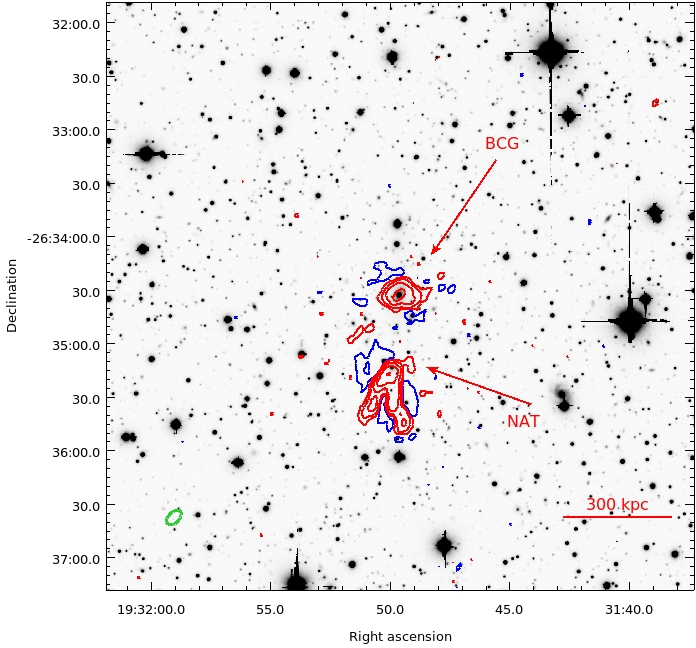}
\caption[The radio-optical overlays]{Left panel: A\,1300 radio contours (red) overlaid on the gray scale Infra-Red band image (K$_s$ band from GCAV). Radio contours start at 3$\sigma$ ($1 \sigma = 0.04$~mJy~beam$^{-1}$) and then scaled by a factor of 2. The first negative contour (-3$\sigma$ level) is drawn in blue. Galaxies are labeled following \citet{Reid99}. Right panel: same as left panel but for MACS\,J1931.8-2634. Radio contours are overlaid on the optical Suprime image (Ic filter). The restoring beam for each image is shown as a green ellipse in the bottom left corner in both panels.}
\label{fig:overlays}
\end{figure*}

Despite the inadequate resolution of our images and the limitations in the $uv$--coverage, the giant radio halo in A\,1300 and the radio relic are visible in the inner region of the field shown in Fig.~\ref{fig:A1300_field}. A few strong sources are distributed over the whole field of view, which is otherwise populated by faint radio sources.
Fig.~\ref{fig:MACSJ1931_field} shows that the field of MACS\,J1931.8--2634 is dominated by faint radio sources, too.
Radio-optical overlays of the central $5.5$~arcmin~$\times$~5.5~arcmin region are displayed in Fig.~\ref{fig:overlays} for both clusters.

The A\,1300 radio galaxies   (Fig.~\ref{fig:overlays}, left panel) are labeled according to \citet{Reid99}.
At the resolution and sensitivity of our images the BCG is associated with a compact radio source, while A2 is most likely a tailed radio galaxy.  Considering the slightly different observing frequencies, the flux density of the sources labeled in Fig.~\ref{fig:overlays} (left panel) are consistent with those in \citet{Reid99}
within the errors. The relic is very clearly imaged, and consistent in size with the 325 MHz image in \citet{Venturi13}. 

The central region of MACS\,J1931.8--2634 (Fig.~\ref{fig:overlays}, right panel) is dominated by the radio emission associated with the BCG and the NAT galaxy. At the angular resolution and sensitivity of our image the morphology of the BCG is fully consistent with the 1.4 GHz VLA--B image reported in \citet{Giacintucci14}, i.e. no further emission is detected surrounding the BCG. The total flux density of the radio emission in our image is $53 \pm 1 $~mJy, slightly lower than reported at 1.4~GHz by \citet{Giacintucci14}, i.e., $S_{1.4 \, {\rm GHz}} = 62 \pm 3$~mJy \citep[see also][]{Ehlert11}. 
The overall morphology and extent of the NAT is very similar to that shown in \citet{Giacintucci14}.

\subsection{Source extraction}

For both clusters we extracted the radio sources using the \texttt{PyBDSF} \citep{Mohan15} package from the images not corrected for the primary beam, where the noise is fairly flat over the whole field. \texttt{PyBDSF} first estimates a noise map by calculating the rms noise over a 16--pixel box sliding by a 50--pixel window. Following \cite{Williams16}, the box was taken about two times smaller around bright sources to account for the possible local noise increase due to calibration artefacts.
The sources were extracted by first identifying islands of contiguous emission above a given threshold, and then decomposing islands into Gaussian components. A threshold of five and four times the local rms noise was used to define sources and  island boundaries respectively. The flux density of the sources was then corrected for primary beam attenuation.
We detected 107 and 162 radio sources within the inner 30~arcmin in the field of A\,1300 and MACS\,J1931.8--2634 respectively above the threshold of 5$\sigma$ (0.2~mJy) in the images not corrected for the primary beam attenuation. Most of the sources detected in both fields are unresolved or barely resolved. We assume that a radio source is resolved when its deconvolved major and minor axes, as given by PyBDSF, are larger than the restoring beam of the image.
The radio source catalogues are reported in Tables~\ref{tab:A1300_Radio_Catalog} and \ref{tab:MACSJ_Radio_Catalog}. 

\begin{figure}
\centering
    \includegraphics[scale=0.55]{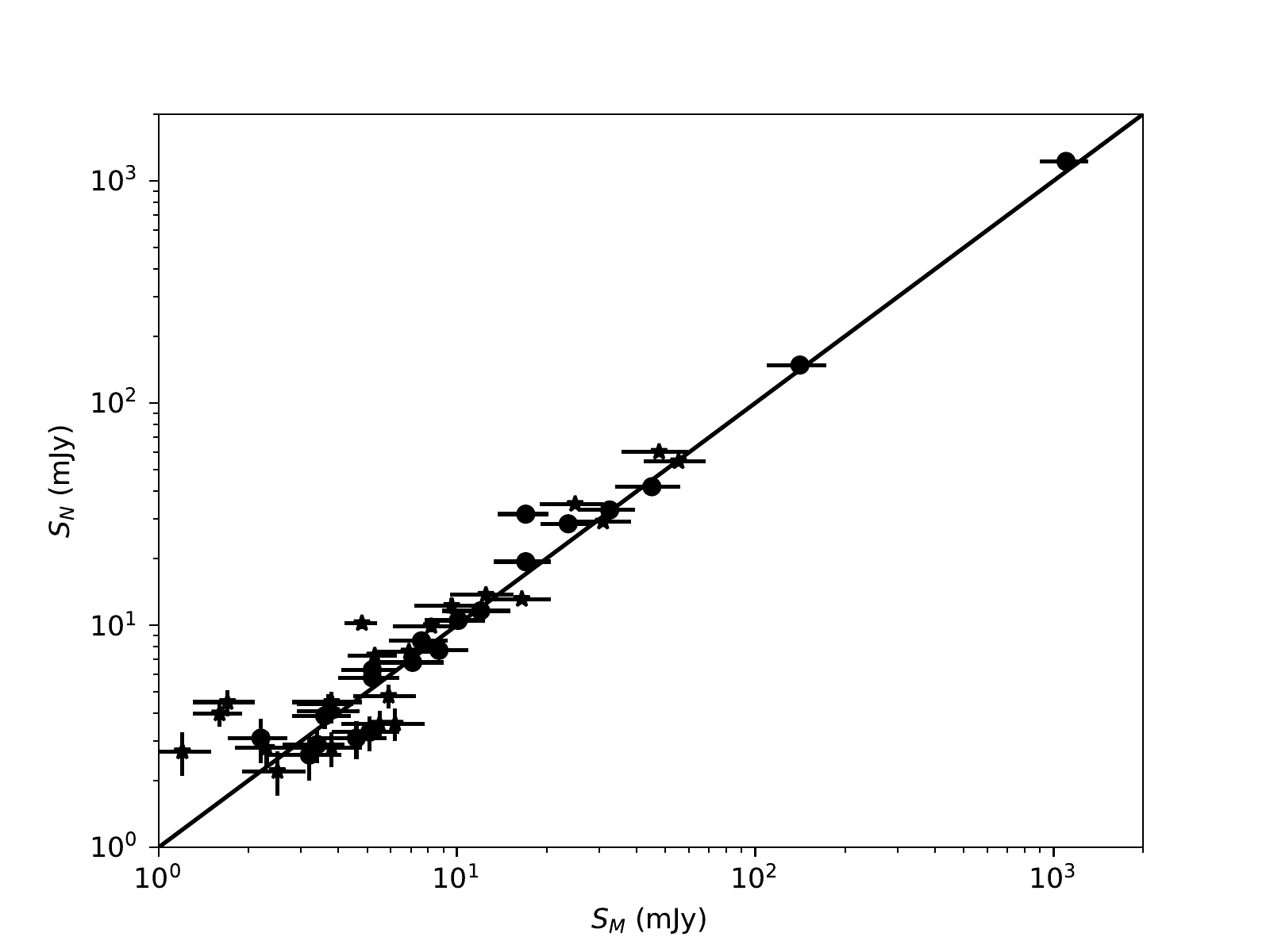}
\caption[The NVSS comparison]{Comparison between the NVSS flux density $S_N$ and the MeerKAT flux density $S_M$ scaled at 1.4 GHz assuming a spectral index $\alpha=0.7$ for sources common to both cluster catalogues within a 30~arcmin radius from the image centre.}
\label{fig:nvss_S14}
\end{figure}

We tested the accuracy of our flux density calibration by comparing our catalogue with the NVSS catalogue \citep{Condon98}. 
We first produced MeerKAT images matching the NVSS 45~arcsec angular resolution, and used them to fit for the flux density for those sources that we considered isolated (i.e. not blended with other sources) and compact (to account for the different angular resolution and $uv$-coverage of MeerKAT--16 and NVSS) within 30~arcmin from the pointing centre.  We finally scaled the
MeerKAT flux density to 1.4~GHz (NVSS observing frequency) assuming a spectral index\footnote{We used the convention $S_\nu \propto \nu^{-\alpha}$ where $S_\nu$ is the flux density at the frequency $\nu$.} $\alpha = 0.7$ for all the sources.
This procedure ensures a one-to-one match between the two catalogues. The result of our comparison is reported in Figure~\ref{fig:nvss_S14}. Although there is some scatter below 10~mJy, the alignment between the two measurements is generally good and the rms of the relative difference is $\sim 16$~per~cent for sources brighter than 8~mJy at 1.28~GHz, which is most likely due to the higher sensitivity of NVSS to extended emission.
We consider this number as an indication of the accuracy of our absolute flux density scale. The errors on the flux density measurements reported in Table~\ref{tab:A1300_Radio_Catalog} and \ref{tab:MACSJ_Radio_Catalog} are the PyBDSF fit errors that do not include the uncertainty on the flux density scale.

\subsection{Radio Source Counts}

As a first step to test possible differences in the radio source population, after primary beam correction we derived the differential radio source counts in a circular area of radius $r = 12$~arcmin.
We chose the same area (0.125~deg$^2$), which corresponds to about two virial radii in both clusters (blue dashed circle in Fig.~\ref{fig:A1300_field} and ~\ref{fig:MACSJ1931_field}). We expect that the cluster dynamics may not impact the radio emission of individual galaxies beyond this distance. 

To account for the different angular resolution of the final images of the two cluster fields (see Table~\ref{tab:Details_observations}) and ensure an homogeneous analysis, we generated an image of the MACS\,J1931.8--2634 field with a 12~arcsec~$\times$~5~arcsec restoring beam, and used this latter image in the following steps.

\begin{figure}
    \centering
    \includegraphics[width=0.49\textwidth]{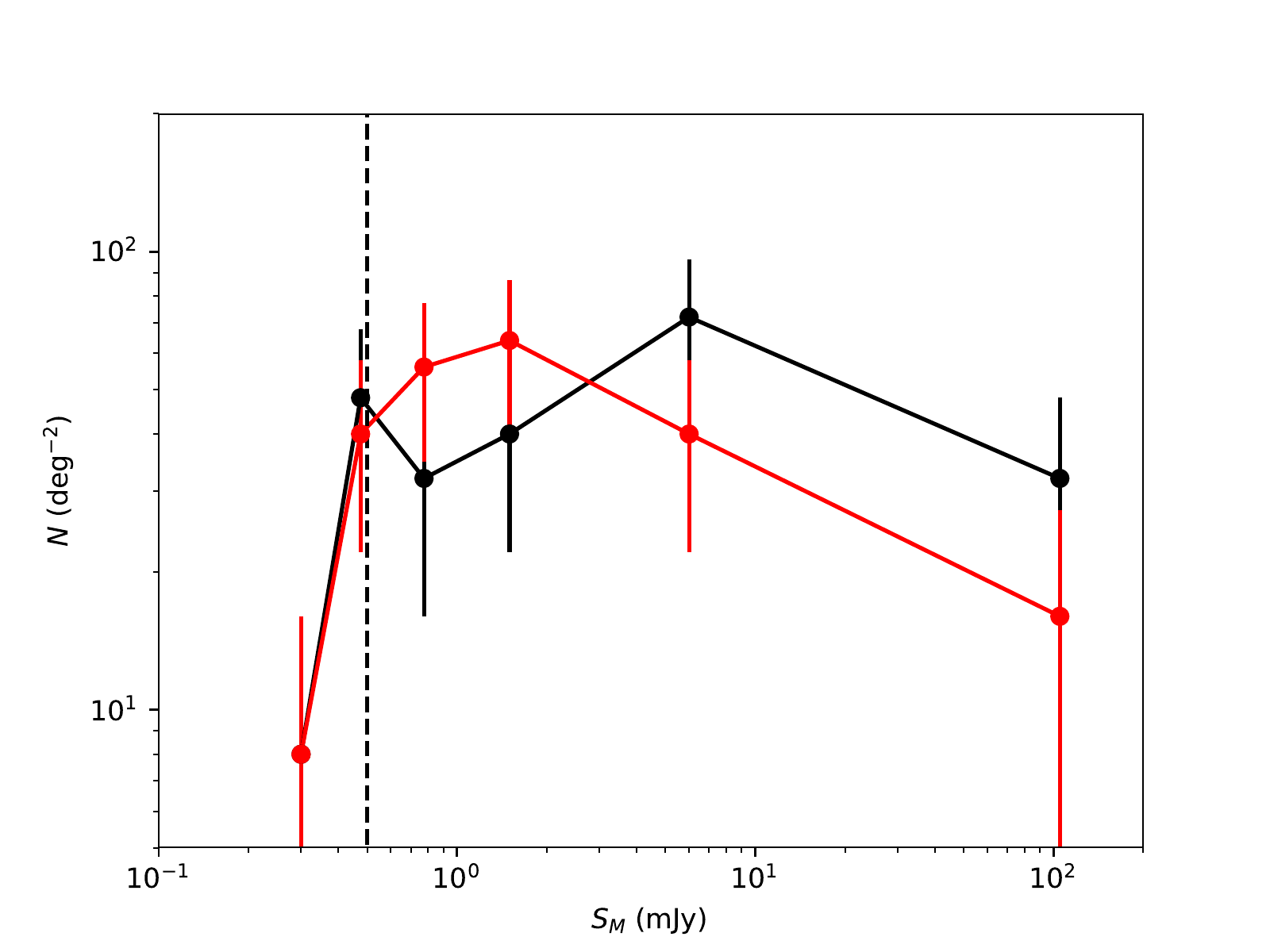}
    \caption{Differential radio source counts in A\,1300 (black) and MACS\,J1931.8-2634 (red) at 1.28~GHz in the central 0.125~deg$^2$ area, corresponding to two virial radii (blue dashed circle in Fig.~\ref{fig:A1300_field} and \ref{fig:MACSJ1931_field}).}
    \label{fig:MACSJ_counts}
\end{figure}

\begin{figure}
    \centering
    \includegraphics[width=0.49\textwidth]{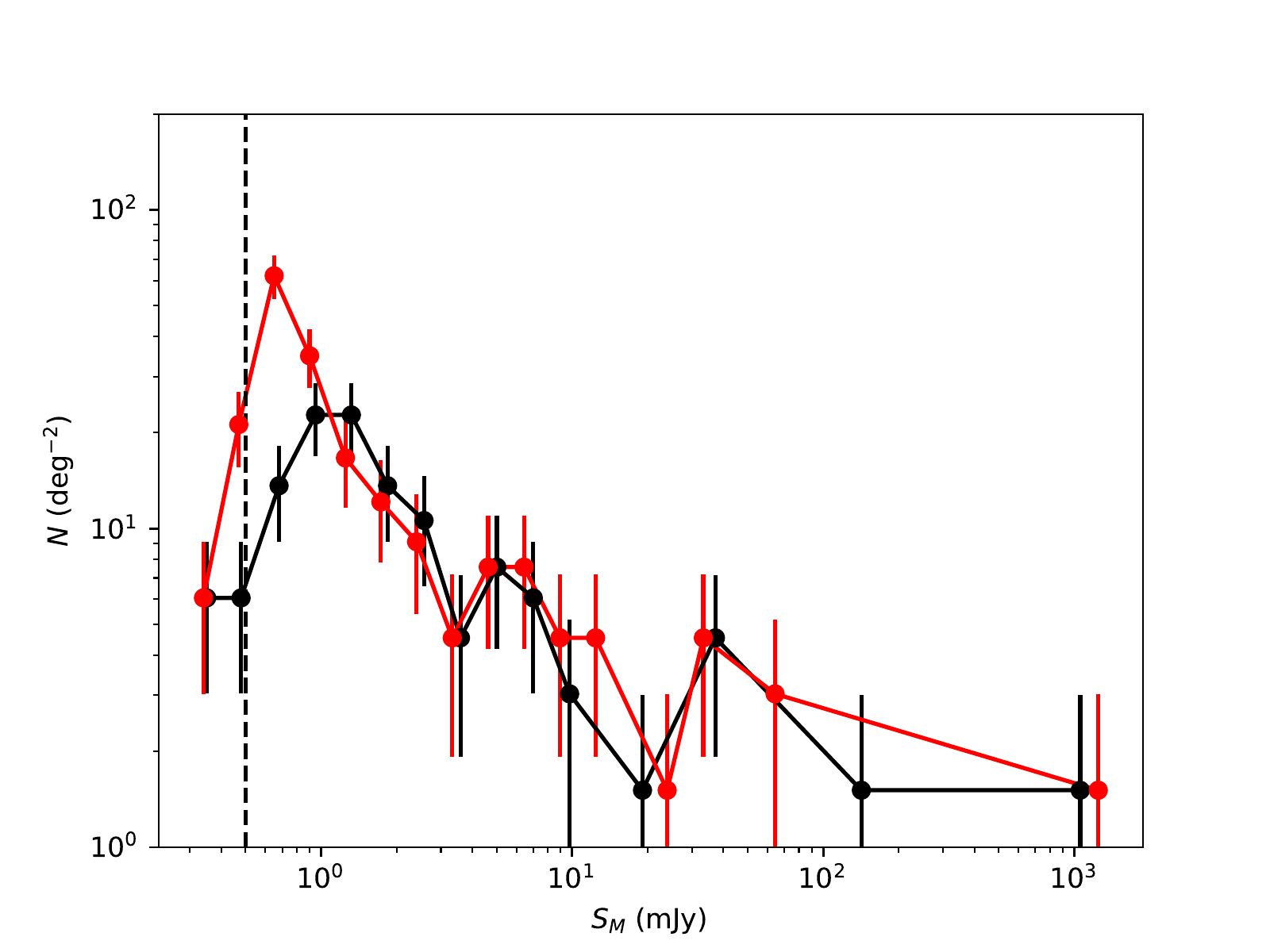}
    \caption{Same as Figure~\ref{fig:MACSJ_counts}, but for the background (the 0.66~deg$^2$ area between the blue and the green circles in Figure~\ref{fig:A1300_field} and \ref{fig:MACSJ1931_field}).}
    \label{fig:A1300_counts}
\end{figure}

To derive the contribution of the two clusters to the differential source counts we estimated the background source counts from the annular region between the blue and green circle (30$^{\prime}$) in Fig.~\ref{fig:A1300_field} and in 
the $12^{\prime\prime}\times5^{\prime\prime}$ image of MACS\,J1931.8--2634, which corresponds to a 0.66~deg$^2$ area.
Fig.~\ref{fig:MACSJ_counts} and \ref{fig:A1300_counts} show the differential source counts $N$ for the 0.125~deg$^2$ area centered on the two clusters and the background (0.66~deg$^2$) respectively. The background counts were not subtracted from 0.125~deg$^2$ central area. The source counts have been normalized to 1 square degree in both figures. 

The source count distributions tend to decline below $\sim 0.5$~mJy, corresponding to $10\sigma$ (dashed line in Fig.~\ref{fig:MACSJ_counts} and \ref{fig:A1300_counts}). We interpret such decline as an indication of the completeness limit of our samples, partly due to the primary beam attenuation, which is $\sim 20$~per~cent at 30~arcmin and negligible at 12~arcmin.
Since our analysis is carried out on images with the same angular resolution, at the same frequencies and, therefore, with the same primary beam, this implies that the completeness is the same for both cases. For the purpose of our relative comparison, the estimate and correction of completeness is therefore not necessary.

We performed a two-sample Kolmogorov-Smirnov (KS) test between the A\,1300 and MACS\,J1931.8--2634 using the source catalogues extracted in their central area (12~arcmin radius). The test yielded a 0.05~$p$-value, implying that we can reject the null hypothesis that the radio source catalogues are drawn from the same distribution with a 95~per~cent significance.
The two distributions for the background source counts are in very good agreement with each other, except for the first two bins at low flux densities.

\subsection{Optical identifications}
\label{sec:optical_identifications}

To search for optical counterparts we used the Rc band for MACS\,J1931.8--2634 and the $r'$ band for A\,1300, the image quality being better since both filters lie redward of the 4000~\AA~break.
We carried out the optical identifications of radio sources in our catalogues within about two virial radii by cross-correlating their positions with the catalogues derived from the  $\it{Subaru}$-SuprimeCam images (see Section 3).
As the primary beam correction is negligible out to this distance from the pointing centre, our optical identifications are complete at the 0.2~mJy flux density level.

The optical position error is $\sigma_o \sim 0.02$~arcsec. We estimated the radio position error $\sigma_r$ as \citep{Prandoni00}:
\begin{equation}
    \sigma_r = \frac{\theta_s}{2 \, {\rm SNR}},
\end{equation}
where $\theta_s$ is the synthesized beam size (Table~\ref{tab:Details_observations}) and SNR is the signal-to-noise ratio. For the faintest sources
(${\rm SNR} = 5\sigma$),  $\sigma_r = 1.2$~arcsec and 0.5~arcsec for A\,1300 and MACS\,J1931.8--2634 respectively.
The matching criterium was based on the $R$ parameter defined as \citep{Giacintucci04}:
\begin{equation}
    R^2= \frac{\Delta^2_{r-o}}{\sigma^2_r + \sigma^2_o} \, ,
\end{equation}
where $\Delta_{r-o}$ is the radio-optical position offset. We considered reliable identifications all the nearest neighbor matches with $R \leq 3$.  
After further visual inspection, we were left with 26 optical counterparts in the A\,1300 cluster field ($\sim 24\%$ of the total number of radio sources), and 25 optical counterparts in the MACS\,J1931.8--2634 cluster field ($\sim 15$~per~cent of the total number of radio sources). 

The information on the optical counterparts was complemented with the redshift data to identify cluster members.
\\
In A\,1300, the redshift information was taken from \citet{Ziparo12}\footnote{ESO GO Large Programme, PI B{\"o}ringer, ID number 169.A--0595}. We retrieved and re-analyzed the VIMOS spectroscopic data in \citet{Ziparo12}, limiting to slit including optical counterparts of radio sources and to targets with $r'$ brighter than 20.4 (see Table~\ref{tab:Details_clusters_members}). Combining our re-analysis with information from NED, we found 14 spectroscopic redshifts, nine of which are consistent with being cluster members. Among these nine galaxies, four are new findings from our re-analysis.

For MACS\,J1931.8--2634, the redshift information was obtained from \citet[private
communication]{Rosati14}\footnote{CLASH VLT-ESO Large Programme, PI Rosati, ID number 186.A-0798}.
We found 8 spectroscopic redshifts, six of which are consistent with being cluster members. 

Even though we did not perform a full pseudo-phase space analysis,  using the rest frame $\Delta (v - v_{cluster})$ and the projected distance from the BCG, assumed to be the center of the cluster, both clusters show that the optical counterparts with spectroscopic redshift lie within the phase-space diagram which envelops the cluster members \citep[Figure 4 in][]{Ziparo12}. Since the two clusters have similar mass, the phase-space diagram for the two clusters is similar, and this has been confirmed from a preliminary analysis of the yet unpublished CLASH-VLT data of MACS\,J1931.8-2634. 
This confirms that the radio galaxies with spectroscopic redshift counterparts in the range 0.29-0.31 and 0.34-0.36 for A\,1300 and MACS\,J1931.8--2634 respectively are most likely cluster members. 

The full list of optical identifications is given in Table~\ref{tab:MACSJ_MKT_crossID} and Table~\ref{tab:A1300_MKT_crossID}, and the radio-optical details of the cluster members 
are given in Table~\ref{tab:Details_clusters_members}.

Although beyond the scope of the current paper, it is worth noting that using SED fitting \citep[MAGPHYS,][]{daCunha08} both in A\,1300 and in MACS\,J1931.8--2634 the cluster radio galaxies split into quiescent and star forming with respect to the specific Star Formation Rate, which reflects the blue and red colours in the colour--magnitude diagrams (not shown here). 
\begin{table*}
\caption[Clusters members]{Details of the clusters members. Left to right: (1) radio source name; (2) and (3) J2000 right ascension and declination; (4) flux density at 1.28~MHz $S_M$; (5) indication on source size (extended or not); (6) $k$-corrected radio power $\rm{P_{1.28~GHz}}$; (7) and (8) J2000 right ascension and declination of the optical counterpart; (9), (10) and (11) magnitudes and corresponding colours; (12) spectroscopic redshift $z_{\rm sp}$.}
\label{tab:Details_clusters_members}
\centering
\begin{footnotesize}
\begin{center}
\begin{tabular}{lr cr cr cr cr cr cr cr cr cr cr cr}
\hline
\\
(1) & (2) & (3) & (4) & (5) & (6) & (7) & (8) &(9) & (10) & (11) & (12) \\
NAME& RA$_{J2000}$& DEC$_{J2000}$& $S_M$ & resolved & log$\rm{P_{1.28~GHz}}$ & RA$_{opt}$& DEC$_{opt}$&   B& Rc & (B-Rc)&  $z_{\rm sp}$&\\
& h m s& $^\circ$ ' "& mJy&  & W~Hz$^{-1}$& hms& $^\circ$ ' "& & & &\\
\hline
J1931-2637 & 19:31:46.60 & -26:37:31.7 & $0.39 \pm 0.04$ & N & 23.22 & 19:31:46.53 & -26:37:31.0 & 21.95 & 19.67 & 2.28 & 0.342\\
J1931-2634 & 19:31:49.58 & -26:34:32.7 & $45.40 \pm 0.80$ & Y--BCG & 25.31 & 19:31:49.63 & -26:34:32.6 & 18.84 & 18.14 & 0.70 & 0.352\\
J1931-2635 & 19:31:50.02 & -26:35:17.2 & $145.20 \pm 0.70$ & Y--NAT & 25.81 & 19:31:50.00 & -26:35:17.1 & 21.50 & 19.20 & 2.30 & 0.351\\
J1931-2630b & 19:31:54.86 & -26:30:57.7 & $0.88 \pm 0.10$ & N & 23.60 & 19:31:54.86 & -26:30:57.0 & 21.66 & 19.46 & 2.20 & 0.351\\
J1931-2645 & 19:31:58.33 & -26:45:34.4 & $0.25 \pm 0.05$ & N & 22.98 & 19:31:58.32 & -26:45:34.4 & 20.85 & 19.33 & 1.52 & 0.359\\
J1932-2625 & 19:32:07.42 & -26:25:16.3 & $0.25 \pm 0.05$ & N & 22.98 & 19:32:07.45 & -26:25:16.5 & 20.52 & 19.22 & 1.30 & 0.349\\
 \hline
 \hline
\\
(1) & (2) & (3) & (4) & (5) & (6) & (7) & (8) &(9) & (10) & (11) & (12) \\
NAME & RA$_{J2000}$ & DEC$_{J2000}$ & $S_M$ & resolved & log$\rm{P_{1.28~GHz}}$ & RA$_{opt}$ & DEC$_{opt}$&   $g'$& $r'$ & $(g'-r')$ &  $z_{\rm sp}$&\\
& h m s & $^\circ$ ' " & mJy &  & W~Hz$^{-1}$ & hms & $^\circ$ ' "& & & & &\\
\hline
 J1131-2000 & 11:31:13.74 & -20:00:20.5 & $1.69 \pm 0.17$ & N & 23.73 & 11:31:13.73 & -20:00:21.5 & 21.35 & 19.95 & 1.40 & 0.303\\
 J1131-1954 & 11:31:46.79 & -19:54:52.5 & $0.25 \pm 0.05$ & N (B2) & 22.89 & 11:31:47.13 & -19:54:52.7 & 21.05 & 19.63 & 1.41 & 0.302\\
 J1131-1949 & 11:31:48.62 & -19:49:01.7 & $0.25 \pm 0.05$ & N & 22.89 & 11:31:48.56 & -19:49:02.2 & 21.23 & 20.21 & 1.02 & 0.302\\
 J1131-1958 & 11:31:49.50 & -19:58:07.9 & $0.39 \pm 0.05$ & N & 23.08 & 11:31:49.53 & -19:58:07.6 & 20.87 & 19.74 & 1.13 & 0.295\\
 J1131-1953b & 11:31:54.30 & -19:53:53.5 & $37.80 \pm 0.40$ & Y--TAIL (A2) & 25.03 & 11:31:54.27 & -19:53:50.8 & 20.86 & 19.41 & 1.45 & 0.305\\
 J1131-1955 & 11:31:54.34 & -19:55:39.0 & $12.10 \pm 0.20$ & Y--BCG (A1)& 24.53 & 11:31:54.18 & -19:55:39.8 & 20.09 & 18.62 & 1.46 & 0.307\\
 J1131-1952b & 11:31:54.85 & -19:52:07.7 & $2.30 \pm 0.20$ & Y (A3) & 23.80 & 11:31:54.95 & -19:52:10.2 & 20.70 & 19.24 & 1.45 & 0.303\\
 J1132-1954 & 11:32:02.77 & -19:54:09.0 & $1.02 \pm 0.10$ & Y (12)& 23.38 & 11:32:02.70 & -19:54:13.5 & 21.23 & 19.78 & 1.44 & 0.306\\
 J1132-1952 & 11:32:04.16 & -19:52:11.1 & $0.25 \pm 0.05$ & N & 22.89 & 11:32:04.39 & -19:52:12.4 & 21.14 & 20.39 & 0.75 & 0.302\\
 \hline
\end{tabular}
\end{center}
\end{footnotesize}
\end{table*}

\subsection{The Radio Luminosity Function}

The RLF is a solid statistical tool to investigate the radio properties of a galaxy population \citep{Ledlow96,Venturi00,Giacintucci04,Branchesi06}.
To further explore if and how the dynamical properties of galaxy clusters affect the statistical properties of the radio galaxy population, we derived the RLF in both clusters. In particular, the influence of the environment should be reflected in the shape of the RLF or in its amplitude, or both.

Consistent with the radio source counts, our analysis has been carried out within about two virial radii (12~arcmin) in both clusters.
The 1.28~GHz $k$-corrected
radio power of the cluster radio galaxies ranges from $22.95 < \log{\rm P_{1.28~GHz}~(W/Hz)} < 25.03$ in A\,1300 and $22.98 < \log{\rm P_{1.28~GHz}~(W/Hz)} < 25.31$ in MAC\,J1931.8--2643 (see Table~\ref{tab:Details_clusters_members}).
The redshift of the individual galaxies was used to evaluate the $k$ correction.
We divided this interval in bins of $\Delta \log P = 0.4$  in A\,1300. Due to the lower number of sources, we chose a bin width of $\Delta \log P = 0.6$ for MACS\,J1931.8--2634.
\\
To ensure optical completeness in the normalization of the radio galaxy distribution we selected 
all the galaxies in the magnitude range $18.6 < m_{r'} < 20.3$ for A\,1300 and those in the magnitude range
$18.1 < m_{\rm Rc} < 19.7$ for MACS\,J1931.8--2634. The negligible impact of the different filters for the two clusters is discussed in Section~\ref{sec:optical}. The faint limits in $r'$
and Rc are approximately 5 magnitudes brighter than the depth of the stacked images, which ensures completeness of the optical samples in both clusters.
We used the $r'$ and Rc filters which are redder than the 4000~\AA \  break for both clusters, and they have similar system transmission as opposed to {\em g}  and B,V filters. 

\begin{figure}
    \centering
    \includegraphics[width=0.48\textwidth]{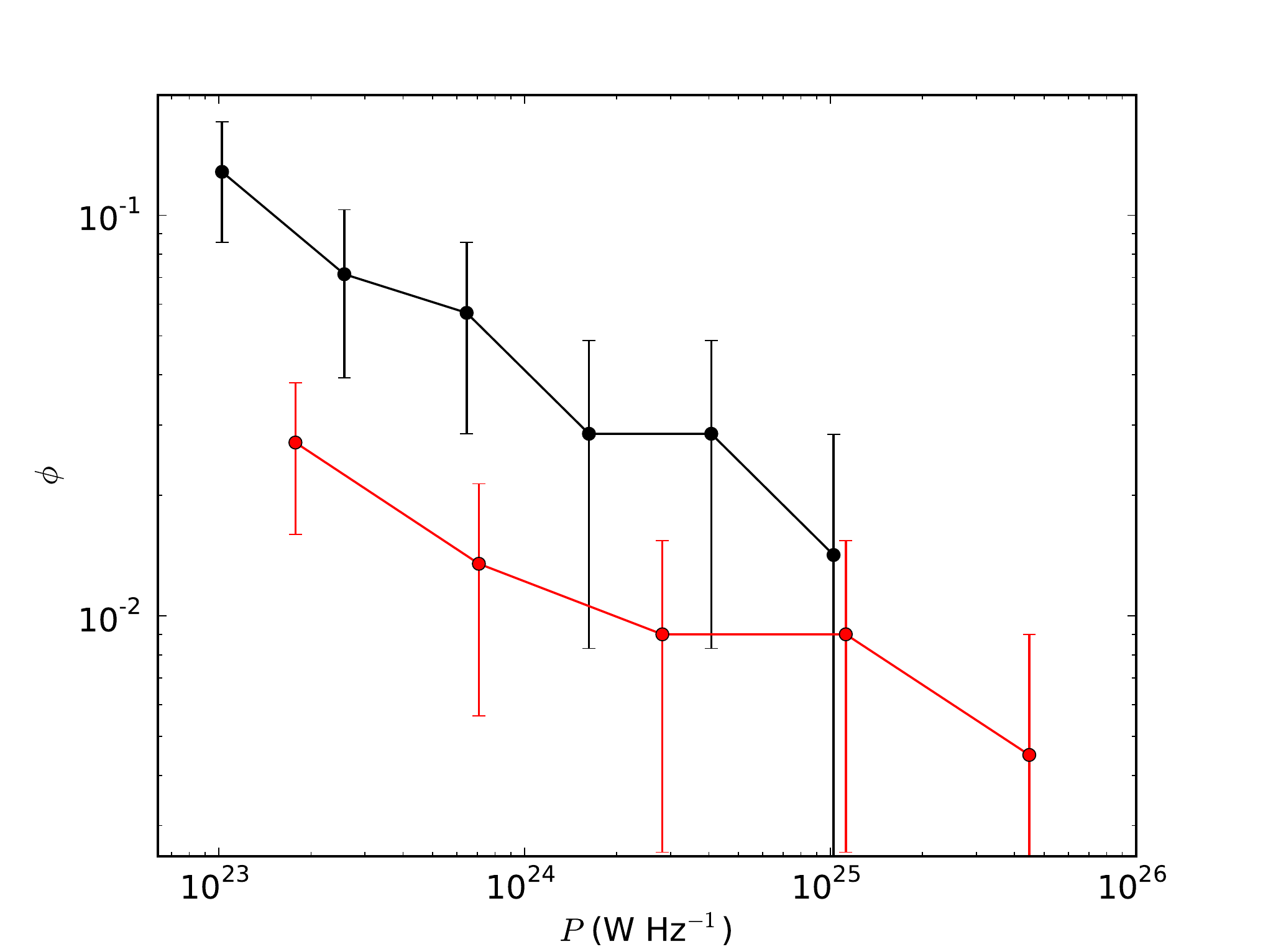}
    \caption{Integral RLF of the A\,1300 (black) and MACS\,J1931.8-2634 (red) cluster, respectively (see text for details).}
    \label{fig:RLF}
\end{figure}

We obtained 261 objects for A\,1300, 70 of which have spectroscopic redshift in the range  $0.29 < z_{\rm sp} < 0.31$, and 1181 objects for MACS\,J1931.8--2634, 220 of which have spectroscopic redshift in the $0.34 < z_{\rm sp} < 0.36$ range. Each radio power bin has been normalized by 70 and 220 respectively for A\,1300 and MACS\,J1931.8--2634, as  clear from Table~\ref{tab:MACSJ_MKT_RLF}, which reports the radio power interval (Col. 1) and the values of the the fractional and integral RLF (Col. 2 and 3 respectively) for both clusters. MACS\,J1931.8--2634 is reported in the upper part of the table,  A\,1300 in the lower part.
The two integral radio luminosity functions are shown in Fig.~\ref{fig:RLF}. 

The integral RLF has a similar slope in both clusters, the one in A\,1300 being systematically higher compared to MACS\,J1931.8--2634 one. Unfortunately,
the samples are not large enough to perform a KS test, and we integrated the RFLs in a single power bin over the full radio power range. 
In order to account for the different radio power detection limit in the two clusters ($\log{\rm P_{1.28~GHz}} = 22.89$ and 22.98 for A\,1300 and MACS\,J1931.8--2634 respectively), we removed the three objects with log${\rm P_{1.28~GHz}} = 22.89$ in A\,1300.
We obtained $0.027 \pm 0.011$ for MACS\,J1931.8-2634 and {\bf $0.09 \pm 0.04$} for A\,1300 respectively, leading to a $3.3 \pm 1.9$ ratio between the two integral RFLs.
These results indicate that the evidence of enhanced radio emission in A\,1300 is negligible ($1.2 \sigma$).

\begin{table}
\caption[MACSJ1931-2634 radio integral LF]{RLFs of MACS\,J1931.8-2634 (top part) and A\,1300 (bottom part) clusters respectively. From left to right: radio power (logarithmic) interval, differential RLF (normalized by the number of cluster optical galaxies) and cumulative RLF.}\label{tab:MACSJ_MKT_RLF}
\begin{footnotesize}
\begin{center}
\begin{tabular}{cr cr cr}
\hline
(1) & (2) & (3) \\
  $\Delta$log${\rm P_{1.28~GHz}}$ & differential RLF & integral RLF &\\
\hline
  22.95-23.55 & 3/220 & 0.0271\\
  23.55-24.15 & 1/220 & 0.0135\\
  24.15-24.75 & 0/220 & 0.009\\
  24.75-25.35 & 1/220 & 0.009\\
  25.35-25.95 & 1/220 & 0.0045\\
\hline
  22.81-23.21 & 4/70 & 0.1284\\
  23.21-23.61 & 1/70 & 0.0713\\
  23.61-24.01 & 2/70 & 0.0571\\
  24.01-24.41 & 0/70 & 0.0285\\
  24.41-24.81 & 1/70 & 0.0285\\
  24.81-25.21 & 1/70 & 0.0142\\
\hline
\end{tabular}
\end{center}
\end{footnotesize}
\end{table}

\section{Summary and future work}\label{sec:discussion}

In this paper we presented new 1.28~GHz MeerKAT observations of the two galaxy clusters A\,1300 and MACS\,J1931.8--2634 as part of the MeerKAT early science programme, when only 16 antennas were available for observing. Observations served to test calibration and imaging pipelines for MeerKAT. As scientific goal, we selected a (A\,1300) and a relaxed cluster (MACS~J\,1931.8--2634) with similar mass located at $z \sim 0.3$ in order to isolate effects due to cosmological evolution, for a comparative study of the properties of the radio galaxy population in different environments. 
A\,1300 is a merging cluster, with a well-studied giant radio halo and a relic, as often found in merging clusters; MACS\,J1931.8--2634 is a relaxed cluster hosting one of the most X--ray luminous cool cores, and extended emission of unclear nature surrounding the radio AGN associated with the BCG.

The angular resolution and $uv$--coverage of our observations are very well suited to perform a comparative study of the statistical properties of the radio galaxy population, which we carried out by means of the radio source counts and the radio luminosity function.

We extracted a radio source catalogue for each cluster down to a 0.2~mJy threshold (corresponding to 5$\sigma$) and out to about two virial radii (corresponding to a 12~arcmin radius) which we consider the extent of the possible effects of cluster dynamics on the radio galaxy population. We further extracted another catalogue from the outskirt of each cluster in an annular area of radius 12~$ < r <$~30~arcmin, representative of the background distribution. 

We performed a two sample KS test of the radio source counts in the two clusters. We rejected the hypothesis that the cluster catalogues come from the same distribution at a limited, 95~per~cent significance.

To further investigate if and how the cluster dynamics affects the radio emission of cluster galaxies, we complemented our radio catalogues with optical data in order to derive the radio-optical luminosity function for each cluster. In particular, we used $Subaru$ SuprimeCam data for A\,1300 in the $g^{\prime}$ and $r^{\prime}$ and the CLASH full filter coverage for MACS\,J1931.8--2634. These datasets are $\approx 5$~magnitudes deeper than the faint limits used for the radio-optical luminosity function.
The optical data were complemented with redshift information from the literature, yielding to the identification of 9 cluster radio galaxies out to 70 cluster members and 6 cluster radio galaxies out of 220 cluster members in A\,1300 and MACS\,J1931.8--2634 respectively. These radio galaxies were used to compute the differential and integral radio luminosity functions which span the power range $22.81 < \log\rm{P_{1.28~GHz}}~(W/Hz) < 25.95$.

We found that the integral RLF in A\,1300 systematically lies above the MACS\,J1931.8--2634 one. After averaging over the whole common power interval, the ratio of the two RLFs is $3.3 \pm 1.9$, implying that the probability to host radio emission is 3.3 times higher in A\,1300. Due to the small sample, however, this result is not statistically significant ($1.2\sigma$ level), suggesting that the role of cluster mergers as a trigger of radio emission in the cluster galaxy population is negligible.

The analysis presented in this work, though limited by the small statistics, aligns with the previous studies in the field, and the role of cluster
dynamics on the global statistical properties of radio galaxies remains unclear (i.e., \citep[i.e.,][]{Ledlow96,Venturi00,Giacintucci04}. 
Future, more sensitive observations with the full MeerKAT array will increase the sample statistics and allow a more robust confirmation of the findings presented here.

\section*{Acknowledgements}
The MeerKAT telescope is operated by the South African Radio Astronomy Observatory, which is a facility of the National Research Foundation, an agency of the Department of Science and Innovation. This work is based on research supported in part by the National Research Foundation of South Africa (Grant Number 103424). We acknowledge the support from the Ministero degli Affari Esteri e della Cooperazione Internazionale, Direzione Generale per la Promozione del Sistema Paese, Progetto di Grande Rilevanza ZA18GR02. Based in part on data collected at Subaru Telescope and obtained from the SMOKA, which is operated by the Astronomy Data Center, National Astronomical Observatory of Japan. Basic research in radio astronomy at the Naval Research Laboratory is supported by 6.1 Base funding. RK acknowledges the support of the Department of Atomic Energy, Government of India, under project no. 12-R\&D-TFR-5.02-0700.

\section*{Data availability}

The data underlying this article will be shared on reasonable request to the corresponding author.




\bibliographystyle{mnras}
\bibliography{biblio} 





\appendix
\onecolumn

\section{Radio Source catalogues}\label{append:cat}

We provide the radio catalogues for A\,1300 (Table~\ref{tab:A1300_Radio_Catalog}) and MACS\,J1931.8--2634 (Table~\ref{tab:MACSJ_Radio_Catalog}) fields extracted within 30' radius from the pointing centre. The \texttt{pyBDSF} fitted positions and MeerKAT flux densities are also indicated.
\begin{footnotesize}
\begin{center}
\begin{longtable}{l cr cr cr cr}
\caption[A1300 Radio Catalogue]{A\,1300 radio source catalogue. The columns report the name of the radio source (Col. 1), J2000 right ascension and declination (Col. 2 and 3 respectively), the total flux density (Col. 4) and a note on the morphology (Col. 5). Radio source names new and follow the J\,HHMM--DDMM nomenclature. All sources with S$_{\rm tot} > $ S$_{\rm peak}$ (where  S$_{\rm tot}$ and  S$_{\rm peak}$ are the total flux density and peak flux density at 1.28~GHz respectively) are considered resolved. Cluster members are listed in boldface.}
\label{tab:A1300_Radio_Catalog} \\
\hline
(1) & (2) & (3) & (4) & (5) \\ 
Name & RA$_\text{J2000}$ & DEC$_\text{J2000}$ & 
  S$_{\rm 1.28~GHz}$ &  Resolved\\
  &    hms         &   $^\circ$ ' "             & 
  mJy                  &           \\
\hline 
J1129--1949 & 11:29:55.82 & -19:49:35.7 &  
$0.60\pm0.14$ &   Y\\
J1129--2004&  11:29:59.64 & -20:04:44.1 &  
27.00 $\pm$ 0.30  & ext. FRII\\   
J1130--1958 & 11:30:05.08 & -19:58:21.9 &  
$0.44\pm0.26$ &   Y\\
J1130--2006 & 11:30:06.61 & -20:06:44.2 &  
$2.07\pm0.11$ &   Y\\
J1130--1939 & 11:30:12.67 & -19:39:19.9 &  
$1.78\pm0.18$ &   Y\\
J1130--2013a & 11:30:21.47 & -20:13:38.3 & 
$6.09\pm0.12$ &   Y\\
J1130--1936 & 11:30:23.66 & -19:36:16.6 &  
$0.43\pm0.32$ &   Y\\
J1130--2009 & 11:30:28.10 & -20:09:27.6 &  
$0.74\pm0.12$ &   Y\\
J1130--1945 & 11:30:28.86 & -19:45:56.9 &  
$0.69\pm0.23$ &   Y\\
J1130--1946 & 11:30:31.16 & -19:46:16.2 &  
$2.32\pm0.12$ &   Y\\
J1130--2007a & 11:30:33.01 & -20:07:45.0 &  
$1.33\pm0.12$ &   Y\\
J1130--1934 & 11:30:35.86 & -19:34:01.6 &  
$2.53\pm0.17$ &   Y\\
J1130--2004a & 11:30:36.33 & -20:04:23.0 &  
$135.1\pm0.4$  &   Y\\
J1130--2007b & 11:30:38.45 & -20:07:35.6 &  
$0.27\pm0.05$ &   Y\\
J1130--1943 & 11:30:46.10 & -19:43:59.4 &  
$1.57\pm0.10$ &   N\\
J1130--2001 & 11:30:50.60 & -20:01:12.1 &  
$1.35\pm0.10$ &   Y\\
J1130--2014 & 11:30:50.75 & -20:14:43.7 &  
$5.75\pm0.09$ &   Y\\
J1130--2008 & 11:30:51.71 & -20:08:55.4 &  
$0.81\pm0.09$ &   Y\\
J1130--2011 & 11:30:52.17 & -20:11:56.7 &  
$2.58\pm0.17$ &   Y\\
J1130--2004b & 11:30:54.40 & -20:04:58.1 & 
$0.97\pm0.07$ &   Y\\
J1130--2013b & 11:30:57.85 & -20:13:27.6 & 
$7.85\pm0.10$ &   Y\\
J1130--1959 & 11:30:58.41 & -19:59:36.6 & 
$2.45\pm0.08$ &   Y\\
J1131--1939 & 11:31:05.15 & -19:39:22.6 &  
$1.61\pm0.10$ &   N\\
J1131--1944a & 11:31:06.20 & -19:44:23.1 & 
$1.13\pm0.13$ &   Y\\
J1131--1954 & 11:31:08.22 & -19:54:17.2 & 
$1.42\pm0.15$ &   Y\\
J1131--1959a & 11:31:09.25 & -19:59:19.0 & 
$4.20\pm0.08$ &   Y\\
J1131--1956 & 11:31:09.94 & -19:56:58.0 &  
$1.49\pm0.08$ &   Y\\
J1131--1944b & 11:31:10.21 & -19:44:42.5 &  
$0.41\pm0.10$ &   Y\\
\textbf{J1131--2000} & \textbf{11:31:13.74} & \textbf{-20:00:20.5} &  
\textbf{1.69$\pm$0.11}&  \textbf{N}\\
J1131--2019a & 11:31:17.81 & -20:19:49.2 &  
$0.57\pm0.20$ &   Y\\
J1131--1952a & 11:31:19.48 & -19:52:39.7 &  
$0.74\pm0.07$ &   Y\\
J1131--1946a & 11:31:20.85 & -19:46:06.7 &  
$0.76\pm0.08$ &   Y\\
J1131--2003 & 11:31:21.65 & -20:03:11.7 &  
$1.47\pm0.07$ &   N\\
J1131--2017 & 11:31:22.36 & -20:17:33.8 &  
$0.29\pm0.16$ &   Y\\
J1131--1947 & 11:31:22.48 & -19:47:31.3 &  
$0.47\pm0.09$ &   Y\\
J1131--2001 & 11:31:24.77 & -20:01:09.6 &  
$10.30\pm0.10$  &   Y\\
J1131--1953a & 11:31:26.89 & -19:53:26.1 &  
$3.91\pm0.08$ &   Y\\
J1131--1951 & 11:31:27.51 & -19:51:46.1 &  
$1.36\pm0.09$ &   N\\
J1131--1931 & 11:31:29.90 & -19:31:48.7 &  
$1.07\pm0.11$ &   Y\\
J1131--1946b & 11:31:32.53 & -19:46:59.5 &  
$0.63\pm0.07$ &   Y\\
J1131--2022 & 11:31:35.60 & -20:22:55.3 &  
$0.47\pm0.19$ &   Y\\
J1131--2007a & 11:31:35.76 & -20:07:28.7 & 
$0.47\pm0.08$ &   Y\\
J1131--1933 & 11:31:39.27 & -19:33:29.6 &  
$10.40\pm0.10$  &   Y\\
J1131--1934 & 11:31:40.32 & -19:34:59.1 &  
$1.05\pm0.12$ &   Y\\
J1131--2019b & 11:31:42.40 & -20:19:25.1 &  
$1.07\pm0.14$ &   Y\\
J1131--1938 & 11:31:43.46 & -19:38:02.8 &  
$1.29\pm0.15$ &   ext. FRII\\
J1131--1952& 11:31:43.92& -19:52:54.2 & 
16.20 $\pm$ 0.10 & ext. (WAT--B1) \\
\textbf{J1131--1954}& \textbf{11:31:46.79}& \textbf{-19:54:52.5}& 
\textbf{0.25$\pm$0.05}& \textbf{N (B2)}&\\
J1131--1950 & 19:31:48.42 & -19:50:43.8& 
$0.70\pm0.08$ & N\\
\textbf{J1131--1949}& \textbf{11:31:48.62}& \textbf{-19:49:01.7}& 
\textbf{0.25$\pm$0.05}& \textbf{N}\\
\textbf{J1131--1958}& \textbf{11:31:49.50}& \textbf{-19:58:07.9}& 
\textbf{0.39$\pm$0.05}& \textbf{N}&\\
J1131-1942 & 11:31:50.85 & -19:42:27.1 &  
$4.78\pm0.21$ &   Y\\
J1131--2012 & 11:31:53.49 & -20:12:11.2 &  
$0.67\pm0.09$ &   Y\\
\textbf{J1131--1953b} & \textbf{11:31:54.30} & \textbf{-19:53:53.5} &  
\textbf{37.80$\pm$0.40}  &  \textbf{ Y--TAIL (A2)}\\
\textbf{J1131--1955}& \textbf{11:31:54.34} & \textbf{-19:55:39.0} &  
\textbf{12.10$\pm$0.20} &  \textbf{Y--BCG (A1)}\\
J1131--1943a & 11:31:54.56 & -19:43:47.6 &  
$0.47\pm0.20$ &   Y\\
\textbf{J1131--1952b} & \textbf{11:31:54.85} & \textbf{-19:52:07.7} &  
\textbf{2.30$\pm$0.13} & \textbf{Y (A3)}\\
J1131-1959b & 11:31:55.60 & -19:59:33.2 &  
$0.40\pm0.17$ &   Y\\
J1131--2008 & 11:31:55.72 & -20:08:18.7 &  
$0.99\pm0.09$ &   Y\\
J1131--2014 & 11:31:55.76 & -20:14:45.5 &  
$1.43\pm0.09$ &   Y\\
J1131--2002 & 11:31:57.47 & -20:02:30.6 &  
$0.35\pm0.15$ &   Y\\
J1131--1946c & 11:31:58.17 & -19:46:56.5 &  
$0.29\pm0.10$  &   Y\\
J1131--2020 & 11:31:58.43 & -20:20:41.1 &  
$0.30\pm0.17$ &   Y\\
J1131--2007b & 11:31:58.69 & -20:07:00.6 &  
$0.40\pm0.14$ &   Y\\
J1131--1943b & 11:31:58.81 & -19:43:53.0 &  
$1.53\pm0.09$ &   Y\\
\textbf{J1132--1954} & \textbf{11:32:02.77} & \textbf{-19:54:09.0} &  
\textbf{1.02$\pm$0.10} &  \textbf{ Y (12)}\\
J1132--1941 & 11:32:03.39 & -19:41:43.1 &  
$0.88\pm0.09$ &   Y\\
\textbf{J1132--1952} & \textbf{11:32:04.16}& \textbf{-19:52:11.1} & 
\textbf{0.25$\pm$0.05}& \textbf{N}&\\
J1132--1926 & 11:32:04.73 & -19:26:17.5 &  
$1.09\pm0.15$ &   Y\\
J1132--2011 & 11:32:06.33 & -20:11:25.9 &  
$1.49\pm0.08$ &   Y\\
J1132--2023 & 11:32:10.98 & -20:23:23.2 &  
$0.88\pm0.15$ &   Y\\
J1132--1956 & 11:32:13.96 & -19:56:09.3 &  
$0.69\pm0.09$ &   Y\\
J1132--2007 & 11:32:14.79 & -20:07:38.9 &  
$0.50\pm0.09$ &   Y\\
J1132--2003a & 11:32:21.68 & -20:03:09.8 &  
$2.19\pm0.08$ &   Y\\
J1132--1951 & 11:32:21.71 & -19:51:41.3 &  
$0.86\pm0.10$  &   Y\\
J1132--2022 & 11:32:22.73 & -20:22:35.3 &  
$7.96\pm0.21$ &   Y\\
J1132--1949a & 11:32:24.85 & -19:49:01.1 &  
$0.50\pm0.19$ &   Y\\
J1132--1930 & 11:32:27.09 & -19:30:35.7 &  
$20.2\pm0.16$&   N\\
J1132--1948 & 11:32:30.09 & -19:48:47.0 &  
$0.72\pm0.08$ &   N\\
J1132--1948b & 11:32:31.26& -19:48:40.0&   
$2.17\pm0.22$ &  Y \\  
J1132--1952 & 11:32:34.77 & -19:52:43.5 &  
$0.25\pm0.11$ &   Y\\
J1132--2022 & 11:32:28.57 & -20:22:54.7 &  
24.00$\pm0.0$ &   Y \\
J1132--2003b & 11:32:37.60 & -20:03:57.8 &  
$1.50\pm0.09$ &   N\\
J1132--1949b & 11:32:37.89 & -19:49:08.7 &  
$6.19\pm0.10$ &   Y\\
J1132--1934 & 11:32:37.94 & -19:34:47.5 &  
$1.81\pm0.11$ &   N\\
J1132--2019 & 11:32:39.35 & -20:19:38.9 &  
$1.52\pm0.11$ &   N\\
J1132--2012 & 11:32:39.67 & -20:12:05.9 &  
$1.16\pm0.10$ &   Y\\
J1132--2006 & 11:32:41.53 & -20:06:54.1 &  
$9.10\pm0.17$ &   Y\\
J1132--2002 & 11:32:43.84 & -20:02:53.7 &  
$4.08\pm0.11$ &   Y\\
J1132--2020 & 11:32:46.00 & -20:20:57.7 &  
$1.11\pm0.14$ &   Y\\
J1132--1957 & 11:32:48.41 & -19:57:55.5 &  
$0.60\pm0.24$ &   Y\\
J1132--2004 & 11:32:59.85 & -20:04:02.7 &  
$7.47\pm0.11$ &   Y\\
J1133--2012 & 11:33:03.79 & -20:12:09.7 &  
$6.70\pm0.14$ &   Y\\
J1133--2018 & 11:33:06.39 & -20:18:55.7 &  
$1.98\pm0.13$ &   N\\
J1133--1952 & 11:33:07.67 & -19:52:04.5 &  
$0.82\pm0.12$ &   Y\\
J1133--1951 & 11:33:08.33 & -19:51:01.7 &  
$0.20\pm0.09$ &   Y\\
J1133--2011 & 11:33:09.56 & -20:11:47.7 &  
$2.50\pm0.14$ &   N\\
J1133--2007 & 11:33:11.27 & -20:07:14.4 &  
$1.05\pm0.15$ &   N\\
J1133--1945 & 11:33:13.05 & -19:45:31.1 &  
$1.24\pm0.11$ &   Y\\
J1133--2004 & 11:33:18.09 & -20:04:24.5 &  
$35.2\pm0.36$ &   Y\\
J1133--2016 & 11:33:20.76 & -20:16:04.0 &  
$1.38\pm0.19$ &   Y\\
J1133--2002 & 11:33:26.29 & -20:02:13.3 &  
$2.61\pm0.26$ &   Y\\
J1133--1940 & 11:33:32.94 & -19:40:22.4 &  
$4.45\pm0.15$ &   Y\\
J1133--1959 & 11:33:33.89 & -19:59:58.5 &  
$4.78\pm0.29$ &   Y\\
J1133--1954 & 11:33:36.91 & -19:54:14.5 &  
$1299.1\pm2.5$&   Y\\
J1133--1959 & 11:33:47.10 & -19:59:05.2 &  
$0.29\pm0.10$ &   Y\\
J1133--1958 & 11:33:57.62 & -19:58:00.0 &  
$0.20\pm0.08$ &   Y\\
\hline 
\end{longtable}
\end{center}
\end{footnotesize}

\begin{footnotesize}
\begin{center}
\begin{longtable}{l cr cr cr cr cr}
\caption[MACSJ Radio Catalogue]{Same as Table~\ref{tab:A1300_Radio_Catalog}, but for MACS\,J1931.8-2634 (cluster members in boldface).}
\label{tab:MACSJ_Radio_Catalog} \\
\hline
(1) & (2) & (3) & (4) & (5) & 
\\
  Name & RA$_\text{J2000}$ & DEC$_\text{J2000}$ &  
  S$_{\rm 1.28~GHz}$ & Resolved\\
  &    hms         &   $^\circ$ ' "           &      
  mJy                       &          \\
\hline
\\
J1929--2627 & 19:29:48.89 & -26:27:32.3 & 
$0.56\pm0.07$ &  Y\\
J1929--2637 & 19:29:53.28 & -26:37:44.7 & 
$2.00\pm0.09$  &  Y\\
J1930--2625a & 19:30:00.09 & -26:25:52.0 & 
$0.30\pm0.04$  &  Y\\
J1930--2631a & 19:30:04.98 & -26:31:46.0 & 
$0.35\pm0.10$  &  Y\\
J1930--2623a & 19:30:05.88 & -26:23:38.5 & 
$0.80\pm0.08$ &  Y\\
J1930--2616a & 19:30:09.21 & -26:16:24.2 & 
$0.54\pm0.13$ &  Y\\
J1930--2640 & 19:30:09.35 & -26:40:42.9 & 
$0.78\pm0.07$ &  Y\\
J1930--2634 & 19:30:09.38 & -26:34:28.8 & 
$0.45\pm0.09$ &  Y\\
J1930--2637 & 19:30:12.67 & -26:37:25.7 & 
$0.26\pm0.11$ &  Y\\
J1930--2650 & 19:30:12.88 & -26:50:49.0 & 
$1.71\pm0.09$ &  Y\\
J1930--2613 & 19:30:13.82 & -26:13:35.5 & 
$0.97\pm0.07$ &  Y\\
J1930--2619 & 19:30:16.26 & -26:19:26.9 & 
$0.80\pm0.07$ &  Y\\
J1930--2644a & 19:30:16.75 & -26:44:30.3 & 
$1.53\pm0.08$ &  Y\\
J1930--2621 & 19:30:16.92 & -26:21:56.6 & 
$0.94\pm0.07$ &  Y\\
J1930--2631b & 19:30:18.81 & -26:31:12.3 & 
$8.29\pm0.13$ &  Y\\
J1930--2624a & 19:30:19.86 & -26:24:11.6 & 
$0.31\pm0.08$ &  Y\\
J1930--2630 & 19:30:21.29 & -26:30:01.3 & 
$2.86\pm0.11$ &  Y\\
J1930--2628 & 19:30:22.94 & -26:28:41.6 & 
$1.13\pm0.07$ &  Y\\
J1930--2643 & 19:30:25.36 & -26:43:40.4 & 
$3.85\pm0.08$ &  Y\\
J1930--2626a & 19:30:25.52 & -26:26:28.7 & 
$2.65\pm0.08$ &  Y\\
J1930--2611 & 19:30:28.42 & -26:11:29.5 & 
$0.25\pm0.12$ &  Y\\
J1930--2626b & 19:30:30.88 & -26:26:24.1 & 
$0.70\pm0.09$ &  Y\\
J1930--2623b & 19:30:37.88 & -26:23:27.0 & 
$0.27\pm0.11$ &  Y\\
J1930--2624b & 19:30:37.95 & -26:24:47.9 & 
$0.49\pm0.09$ &  Y\\
J1930--2643& 19:30:38.23& -26:43:11.4& 
27.10 $\pm$ 0.30& ext. FRII\\
J1930--2625b & 19:30:38.38 & -26:25:06.6 & 
$0.38\pm0.11$ &  Y\\
J1930--2647a & 19:30:38.45 & -26:47:16.1 & 
$2.13\pm0.14$ &  Y\\
J1930--2629 & 19:30:39.43 & -26:29:13.6 & 
$0.75\pm0.08$ &  Y\\
J1930--2653 & 19:30:43.65 & -26:53:45.9 & 
$0.84\pm0.09$ &  Y\\
J1930--2647b & 19:30:44.07 & -26:47:34.8 & 
$0.93\pm0.09$ &  Y\\
J1930--2628 & 19:30:45.91 & -26:28:23.1 & 
$0.58\pm0.10$ &  Y\\
J1930--2644b & 19:30:47.62 & -26:44:15.3 & 
$54.80\pm0.70$  &  Y\\
J1930--2633 & 19:30:48.46 & -26:33:39.8 & 
$0.69\pm0.08$ &  Y\\
J1930--2631c & 19:30:50.58 & -26:31:52.8 & 
$0.71\pm0.10$  &  Y\\
J1930--2658 & 19:30:53.09 & -26:58:00.7 & 
$0.29\pm0.04$ &  Y\\
J1930--2610 & 19:30:53.48 & -26:10:37.6 & 
$0.29\pm0.12$ &  Y\\
J1930--2650 & 19:30:54.68 & -26:50:50.5 & 
$2.80\pm0.08$   &  Y\\
J1930--2655 & 19:30:55.04 & -26:55:37.8 & 
$0.72\pm0.08$ &  Y\\
J1930--2620 & 19:30:55.36 & -26:20:48.4 & 
$12.60\pm0.20$  &  Y\\
J1930--2616b & 19:30:56.61 & -26:16:56.6 & 
$2.00\pm0.08$   &  Y\\
J1930--2636 & 19:30:57.98 & -26:36:14.6 & 
$1.70\pm0.08$   &  Y\\
J1930--2617 & 19:30:59.18 & -26:17:33.9 & 
$0.47\pm0.10$  &  Y\\
J1930--2626c & 19:30:59.84 & -26:26:40.8 & 
$0.30\pm0.08$  &  Y\\
J1931--2616 & 19:31:00.24 & -26:16:45.8 & 
$0.43\pm0.10$  &  Y\\
J1931--2607 & 19:31:00.29 & -26:07:41.4 & 
$1.04\pm0.07$ &  Y\\
J1931--2629 & 19:31:03.06 & -26:29:00.8 & 
$0.40\pm0.08$ &  Y\\
J1931--2620 & 19:31:03.58 & -26:20:25.0 & 
$7.40\pm0.10$    &  Y\\
J1931--2656a & 19:31:05.65 & -26:56:40.6 & 
$0.51\pm0.1$  &  Y\\
J1931--2608a & 19:31:06.99 & -26:08:09.2 & 
$0.29\pm0.12$ &  Y\\
J1931--2640 & 19:31:07.24 & -26:40:59.0 & 
$0.93\pm0.08$ &  Y\\
J1931--2613a & 19:31:09.22 & -26:13:04.0 & 
$0.30\pm0.13$ &  Y\\
J1931--2644 & 19:31:17.19 & -26:44:48.9 & 
$28.40\pm0.30$  &  Y\\
J1931--2608b & 19:31:17.54 & -26:08:44.5 & 
$1.20\pm0.07$  &  Y\\
J1931--2630 & 19:31:18.34 & -26:30:18.4& 
$0.30\pm0.05$& N \\
J1931--2630a & 19:31:19.36 & -26:30:17.4 & 
$0.38\pm0.11$ &  Y\\
J1931--2656b & 19:31:20.25 & -26:56:29.7 & 
$0.67\pm0.09$ &  Y\\
J1931--2640 & 19:31:20.53 & -26:40:46.9 & 
$0.59\pm0.09$ &  Y\\
J1931--2659 & 19:31:27.54 & -26:59:30.7 & 
$1.47\pm0.14$ &  Y\\
J1931--2618 & 19:31:28.04 & -26:18:46.2 & 
$1.10\pm0.08$ &  Y\\
J1931--2651 & 19:31:30.40 & -26:51:23.5 & 
$0.41\pm0.13$ &  Y\\
J1931--2619 & 19:31:30.62 & -26:19:58.6 & 
$11.1\pm0.18$ &  Y\\
J1931--2655 & 19:31:32.41 & -26:55:45.3 & 
$3.71\pm0.08$ &  Y\\
J1931--2654a & 19:31:35.79 & -26:54:52.3 & 
$0.39\pm0.11$ &  Y\\
J1931--2633 & 19:31:36.44 & -26:33:09.6 & 
$14.5\pm0.13$ &  Y\\
J1931--2613b & 19:31:38.92 & -26:13:13.6 & 
$1.50\pm0.08$ &  Y\\
J1931--2617 & 19:31:43.67 & -26:17:42.2 & 
$1.17\pm0.07$ &  Y\\
J1931--2654b & 19:31:45.51 & -26:54:24.2 & 
$0.23\pm0.10$ &  Y\\
\textbf{J1931--2637} & \textbf{19:31:46.60} & \textbf{-26:37:31.7} & 
\textbf{0.39$\pm$0.04} &  \textbf{N}\\
\textbf{J1931--2634} & \textbf{19:31:49.58} & \textbf{-26:34:32.7}& 
\textbf{45.40$\pm$0.80}& \textbf{ext. BCG}&\\
\textbf{J1931--2635} & \textbf{19:31:50.02} & \textbf{-26:35:17.2} & 
\textbf{145.20$\pm$0.70}&\textbf{ext. NAT}\\
J1931--2656 & 19:31:50.81 & -26:56:11.6 & 
$0.50\pm0.14$ &  Y\\
J1931--2624 & 19:31:51.96 & -26:24:27.0 & 
$0.84\pm0.09$ &  Y\\
\textbf{J1931--2630b} & \textbf{19:31:54.86} & \textbf{-26:30:57.7} & 
\textbf{0.88$\pm$0.10}  & \textbf{N}\\
J1931--2622a & 19:31:57.36 & -26:22:09.6 & 
$0.79\pm0.1$ &  Y\\
J1931--2645a & 19:31:57.54& -26:45:37.4& 
$0.40\pm0.05$ & N\\
\textbf{J1931--2645} & \textbf{19:31:58.33} & \textbf{-26:45:34.4} & 
\textbf{0.25$\pm$0.05} & \textbf{N}\\
J1931--2622b & 19:31:58.65 & -26:22:27.5 & 
$0.49\pm0.08$ &  Y\\
J1931--2638 & 19:31:58.92 & -26:38:15.7 & 
$3.40\pm0.09$   &  Y\\
J1932--2654a & 19:32:00.42 & -26:54:17.6 & 
$1.08\pm0.09$ &   Y\\
J1932--2620 & 19:32:03.09 & -26:20:11.3 & 
$0.59\pm0.16$ &  Y\\
J1932--2632 & 19:32:05.17 & -26:32:05.5 & 
$0.48\pm0.11$ &  Y\\
J1932--2627a & 19:32:05.18 & -26:27:00.2 & 
$2.29\pm0.13$ &   Y\\
J1932--2646 & 19:32:06.44 & -26:46:31.2 & 
$0.48\pm0.09$ &  Y\\
J1932--2621a & 19:32:07.11 & -26:21:05.1 & 
$0.60\pm0.09$ &  Y\\
\textbf{J1932--2625} & \textbf{19:32:07.42} & \textbf{-26:25:16.3} & 
\textbf{0.25$\pm$0.05} & \textbf{N}\\
J1932--2702 & 19:32:07.72 & -27:02:28.1 & 
$0.33\pm0.07$ &  Y\\
J1932--2627b & 19:32:08.17 & -26:27:03.6 & 
$0.52\pm0.10$  &  Y\\
J1931--2625 &  19:32:09.72 & -26:25:29.6& 
$0.30\pm0.05$& N\\
J1932--2621b & 19:32:10.80 & -26:21:50.4 & 
$1.74\pm0.12$ &  Y\\
J1932--2634 & 19:32:11.31 & -26:34:01.2 & 
$0.32\pm0.09$ &  Y\\
J1932--2657a & 19:32:11.48 & -26:57:08.6 & 
$1.16\pm0.08$ &  Y\\
J1932--2618a & 19:32:11.85 & -26:18:43.7 & 
$2.20\pm0.13$   &  Y\\
J1932--2637 & 19:32:12.16 & -26:37:30.1 & 
$0.33\pm0.07$ &  Y\\
J1932--2648a & 19:32:12.23 & -26:48:18.1 & 
$0.69\pm0.07$ &  Y\\
J1932--2633a & 19:32:12.53 & -26:33:14.4 & 
$0.90\pm0.09$  &  Y\\
J1932--2628a & 19:32:12.95 & -26:28:07.8 & 
$0.32\pm0.08$ &  Y\\
J1932--2654b & 19:32:14.00 & -26:54:24.0 & 
$0.47\pm0.12$ &  Y\\
J1932--2651 & 19:32:14.41 & -26:51:39.6 & 
$0.47\pm0.13$ &  Y\\
J1932--2648b & 19:32:16.55 & -26:48:44.9 & 
$0.40\pm0.10$ &  Y\\
J1932--2608 & 19:32:16.77 & -26:08:22.2 & 
$4.30\pm0.07$   &  Y\\
J1932--2629 & 19:32:17.70 & -26:29:32.1 & 
$1.50\pm0.09$   &  Y\\
J1932--2621c & 19:32:18.84 & -26:21:26.1 & 
$0.30\pm0.09$   &  Y\\
J1932--2618b & 19:32:19.56 & -26:18:15.1 & 
$0.44\pm0.09$ &  Y\\
J1932--2631 & 19:32:20.07 & -26:31:35.2 & 
$0.50\pm0.06$ & N\\
J1932--2659 & 19:32:21.12 & -26:59:02.7 & 
$0.39\pm0.13$ &  Y\\
J1932--2644 & 19:32:21.41 & -26:44:20.6 & 
$0.74\pm0.14$ &  Y\\
J1932--2654c & 19:32:25.41 & -26:54:26.0 & 
$2.10\pm0.09$   &  Y\\
J1932--2648c & 19:32:25.80 & -26:48:16.4 & 
$0.79\pm0.09$ &  Y\\
J1932--2632 & 19:32:26.13&  -26:32:48.8 & 
$0.50\pm0.07$& N\\
J1932--2633b & 19:32:27.31 & -26:33:10.4 & 
$17.40\pm0.20$  &  Y\\
J1932--2613 & 19:32:29.01 & -26:13:24.3 & 
$0.44\pm0.09$ &  Y\\
J1932--2628b & 19:32:30.33 & -26:28:10.2 & 
$0.43\pm0.10$ &  Y\\
J1932--2628c & 19:32:30.69 & -26:28:50.9 & 
$0.60\pm0.11$ &  Y\\
J1932--2624 & 19:32:30.76 & -26:24:35.8 & 
$0.42\pm0.10$ &  Y\\
J1932--2628d & 19:32:30.92 & -26:28:03.3 & 
$0.85\pm0.11$ &  Y\\
J1932--2641 & 19:32:31.01 & -26:41:51.4 & 
$0.56\pm0.08$ &  Y\\
J1932--2631a & 19:32:31.57 & -26:31:20.8 & 
$0.50\pm0.06$& N\\ 
J1932--2627c & 19:32:31.76 & -26:27:58.9 & 
$1.10\pm0.18$ &  Y\\
J1932--2654d & 19:32:31.78 & -26:54:36.2 & 
$0.65\pm0.09$ &  Y\\
J1932--2639a & 19:32:33.34 & -26:39:08.1 & 
$0.60\pm0.10$ &  Y\\
J1932--2617 & 19:32:36.58 & -26:17:11.1 & 
$0.29\pm0.18$ &  Y\\
J1932--2618c & 19:32:40.26 & -26:18:54.8 & 
$6.40\pm0.09$    &  Y\\
J1932--2645 & 19:32:40.28 & -26:45:47.1 & 
$0.47\pm0.07$   &  Y\\
J1932--2658 & 19:32:41.09 & -26:58:37.8 & 
$3.50\pm0.10$    &  Y\\
J1932--2635 & 19:32:41.52 & -26:35:58.4 & 
$1.23\pm0.11$  &  Y\\
J1932--2657b & 19:32:43.01 & -26:57:54.5 & 
$0.85\pm0.08$  &  Y\\
J1932--2630 & 19:32:44.91 & -26:30:13.3 & 
$57.80\pm0.30$   &  Y\\
J1932--2639b & 19:32:44.93 & -26:39:32.5 & 
$0.43\pm0.11$  &  Y\\
J1932--2638a & 19:32:46.69 & -26:38:52.2 & 
$0.42\pm0.11$  &  Y\\
J1932--2638b & 19:32:47.75 & -26:38:47.3 & 
$0.83\pm0.08$  &  N\\
J1932--2656 & 19:32:49.35 & -26:56:06.7 & 
$0.49\pm0.22$  &  Y\\
J1932--2642 & 19:32:51.32 & -26:42:51.1 & 
$6.10\pm0.09$    &  Y\\
J1932--2616 & 19:32:57.12 & -26:16:41.9 & 
$0.60\pm0.08$  &  Y\\
J1932--2636 & 19:32:58.43 & -26:36:50.5 & 
$1.60\pm0.09$   &  Y\\
J1932--2643 & 19:32:58.68 & -26:43:29.7 & 
$5.30\pm0.09$    &  Y\\
J1933--2618 & 19:33:04.09 & -26:18:01.8 & 
$0.71\pm0.15$  &  Y\\
J1933--2617 & 19:33:04.44 & -26:17:05.4 & 
$0.45\pm0.10$  &  Y\\
J1933--2630 & 19:33:05.31 & -26:30:04.7 & 
$1.47\pm0.09$   &  Y\\
J1933--2658 & 19:33:05.51 & -26:58:20.8 & 
$5.30\pm0.20$     &  Y\\
J1933--2622 & 19:33:06.31 & -26:22:22.9 & 
$12.3\pm0.1$    &  Y\\
J1933--2652 & 19:33:06.37 & -26:52:35.7 & 
$0.80\pm0.11$  &  Y\\
J1933--2656 & 19:33:10.77 & -26:56:26.2 & 
$0.93\pm0.10$  &  Y\\
J1933--2649 & 19:33:12.35 & -26:49:15.0 & 
$0.31\pm0.08$  &  Y\\
J1933--2613 & 19:33:14.23 & -26:13:22.9 & 
$27.60\pm0.20$    &  Y\\
J1933--2624a & 19:33:17.35 & -26:24:15.5 & 
$0.50\pm0.05$  &  Y\\
J1933--2631 & 19:33:19.13 & -26:31:30.7 & 
$1.24\pm0.11$   &  Y\\
J1933--2640 & 19:33:19.22 & -26:40:31.4 & 
$6.00\pm0.11$    &  Y\\
J1933--2624b & 19:33:23.12 & -26:24:02.8 & 
$0.81\pm0.11$  &  Y\\
J1933--2614 & 19:33:27.28 & -26:14:53.9 & 
$1.30\pm0.09$    &  Y\\
J1933--2634 & 19:33:27.33 & -26:34:04.0 & 
$1.40\pm0.14$   &  Y\\
J1933--2626 & 19:33:28.07 & -26:26:00.8 & 
$1.64\pm0.09$   &  Y\\
J1933--2650 & 19:33:28.26 & -26:50:03.9 & 
$0.37\pm0.11$   &  Y\\
J1933--2618 & 19:33:28.99 & -26:18:34.3 & 
$1.13\pm0.09$  &  Y\\
J1933--2629 & 19:33:30.66 & -26:29:14.8 & 
$6.02\pm0.18$  &  Y\\
J1933--2652 & 19:33:31.35 & -26:52:16.8 & 
$0.69\pm0.14$  &  Y\\
J1933--2642 & 19:33:35.87 & -26:42:44.8 & 
$1.12\pm0.10$  &  Y\\
J1933--2643 & 19:33:38.55 & -26:43:13.9 & 
$0.55\pm0.08$  &  Y\\
J1933--2646 & 19:33:39.12 & -26:46:23.9 & 
$0.77\pm0.09$  &  Y\\
J1933--2623 & 19:33:45.81 & -26:23:41.1 & 
$0.83\pm0.13$  &  Y\\
J1933--2633 & 19:33:48.45 & -26:33:16.4 & 
$748.4\pm6.1$  &  ext. FRII\\
J1933--2645a & 19:33:52.33 & -26:45:33.1 & 
$4.50\pm0.12  $  &  Y\\
J1933--2645b & 19:33:52.53 & -26:45:00.2 & 
$29.30\pm0.20 $  &  Y\\
\\
\hline
\end{longtable}
\end{center}
\end{footnotesize}

\newpage
\section{Radio-Optical cross ID}\label{append:crossmatch}
The radio optical crossmatches for MACS\,J1931.8--2634 and A\,1300 are given in Tables~\ref{tab:MACSJ_MKT_crossID} and \ref{tab:A1300_MKT_crossID} respectively. 
\begin{table}
\caption[MACSJ1931 cross ID catalogue]{MACSJ1931.8--2634 radio-optical identifications (cluster members are in boldface). Left to right: radio source name; J2000 right ascension and declination from the radio image;  J2000 right ascension and declination of the optical counterpart; B and Rc magnitudes; colour index, spectroscopic redshift and MeerKAT flux density.}
\label{tab:MACSJ_MKT_crossID}
\begin{footnotesize}
\begin{center}
\begin{tabular}{cr cr cr cr cr cr cr cr cr cr}
\hline
\\
(1) & (2) & (3) & (4) & (5) & (6) & (7) & (8) & (9) & (10)\\
  Name     & RA$_\text{J2000}$& DEC$_\text{J2000}$& RA$_{opt}$& DEC$_{opt}$& B& Rc& B-Rc& z$_\text{spec}$& S$_{M}$\\               
           &    hms         &   $^\circ$ ' "           &    hms &     $^\circ$ ' " &           &      &   &   & mJy&\\
\hline
\\
  J1930--2636 & 19:30:57.98 & -26:36:14.6 & 19:30:57.97 & -26:36:14.4 & 25.6 & 23.86 & 1.74 & -& $1.70\pm0.08$\\
  J1931--2629 & 19:31:03.06 & -26:29:00.8 & 19:31:03.04 & -26:28:59.8 & 25.45 & 23.58 & 1.87 & -& $0.40\pm0.08$\\
  J1931--2640 & 19:31:07.24 & -26:40:59.0 & 19:31:07.24 & -26:40:58.5 & 25.6 & 23.95 & 1.65 & -& $0.93\pm0.08$\\
  J1931--2630 & 19:31:18.34 & -26:30:18.4 & 19:31:18.30 & -26:30:17.6 & 22.36 & 22.18 & 0.18 & -& $0.30\pm0.05$\\
  J1931--2630 & 19:31:19.36 & -26:30:17.4 & 19:31:19.33 & -26:30:16.9 & 24.65 & 22.94 & 1.71 & -& 0.$38\pm0.11$\\
  J1931--2640 & 19:31:20.53 & -26:40:46.9 & 19:31:20.51 & -26:40:46.1 & 18.84 & 18.78 & 0.06 & -& $0.59\pm0.09$\\
  J1931--2633 & 19:31:36.44 & -26:33:09.6 & 19:31:36.43 & -26:33:09.1 & 24.21 & 21.90 & 2.31 & -& $14.50\pm0.13$\\
  {\bf J1931--2637} & {\bf 19:31:46.60} & {\bf -26:37:31.7} & {\bf 19:31:46.53} & {\bf -26:37:31.0} & {\bf 21.95} & {\bf 19.67} & {\bf 2.28} & {\bf 0.342} & $0.39\pm0.04$\\
  \bf{J1931--2634} & \bf{19:31:49.58} & \bf{-26:34:32.7} & \bf{19:31:49.63} & \bf{-26:34:32.6} & \bf{18.84} & \bf{18.14} & \bf{0.70} & \bf{0.352} & $45.40\pm0.80$\\
  \bf{J1931--2635} & \bf{19:31:50.02} & \bf{-26:35:17.2} & \bf{19:31:50.00} & \bf{-26:35:17.1} & \bf{21.50} & \bf{19.20} & \bf{2.27} & \bf{0.351} & $145.2\pm0.7$\\
  \bf{J1931--2630b} & \bf{19:31:54.86} & \bf{-26:30:57.7} & \bf{19:31:54.86} & \bf{-26:30:57.0} & \bf{21.66} & \bf{19.46} & \bf{2.20} & \bf{0.351} & $0.88\pm0.10$\\
  J1931--2645 & 19:31:57.54 & -26:45:37.4 & 19:31:57.55 & -26:45:37.4 & 24.37 & 22.92 & 1.45 & -&$0.40\pm0.05$\\
  \bf{J1931--2645} & \bf{19:31:58.33} & \bf{-26:45:34.4} & \bf{19:31:58.32} & \bf{-26:45:34.4} & \bf{20.85} & \bf{19.33} & \bf{1.52} & \bf{0.359} & $0.25\pm0.05$\\
  J1931--2638 & 19:31:58.92 & -26:38:15.7 & 19:31:58.93 & -26:38:15.2 & 23.50 & 22.08 & 1.42 & -& $3.40\pm0.09$\\
  J1932--2632 & 19:32:05.17 & -26:32:05.5 & 19:32:05.24 & -26:32:05.8 & 24.49 & 24.41 & 0.08 & -& $0.48\pm0.11$\\
  \bf{J1932--2625} & \bf{19:32:07.42} & \bf{-26:25:16.4} & \bf{19:32:07.45} & \bf{-26:25:16.3} & \bf{20.52} & \bf{19.22} & \bf{1.30} & \bf{0.349} & $0.25\pm0.05$\\
  J1932--2625 & 19:32:09.72 & -26:25:29.6 & 19:32:09.74 & -26:25:28.8 & 22.61 & 21.47 & 1.14 & -& $0.30\pm0.05$\\
  J1932--2637 & 19:32:12.16 & -26:37:30.1 & 19:32:12.11 & -26:37:29.5 & 20.07 & 18.81 & 1.26 & 0.207&$0.33\pm0.07$\\
  J1932--2629 & 19:32:17.70 & -26:29:32.1 & 19:32:17.70 & -26:29:31.6 & 22.33 & 20.31 & 2.02 & -& $1.50\pm0.09$\\
  J1932--2631 & 19:32:20.07 & -26:31:35.2 & 19:32:20.04 & -26:31:34.4 & 19.90 & 18.86 & 1.04 & 0.145& $0.50\pm0.07$\\
  J1932--2632 & 19:32:26.13 & -26:32:48.8 & 19:32:26.13 & -26:32:48.4 & 21.80 & 20.22 & 1.58 & -& $0.50\pm0.07$\\
  J1932--2633 & 19:32:27.71 & -26:33:56.4 & 19:32:27.77 & -26:33:55.6 & 25.6 & 24.16 & 1.44 & -& $17.40\pm0.20$\\
  J1932--2628 & 19:32:30.92 & -26:28:03.3 & 19:32:30.91 & -26:28:02.8 & 25.98 & 22.33 & 3.65 & -& $0.85\pm0.11$\\
  J1932--2641 & 19:32:31.01 & -26:41:51.4 & 19:32:31.04 & -26:41:51.1 & 25.6 & 24.20 & 1.4 & -& $0.56\pm0.08$\\
  J1932--2631 & 19:32:31.57 & -26:31:20.8 & 19:32:31.61 & -26:31:20.4 & 24.24 & 23.49 & 0.75 & -& $0.50\pm0.06$\\
  \\
\hline
\end{tabular}
\end{center}
\end{footnotesize}
\end{table}

\begin{table}
\caption[A1300 cross ID catalogue]{A1300 radio-optical identifications (cluster members in boldface. Left to right: radio source name; J2000 right ascension and declination from the radio image;  J2000 right ascension and declination of the optical counterpart; \textit{g'} and \textit{r'} magnitudes; colour index; spectroscopic redshift and MeerKAT flux density).}\label{tab:A1300_MKT_crossID}
\begin{footnotesize}
\begin{center}
\begin{tabular}{cr cr cr cr cr cr cr cr cr cr}
\hline
\\
(1) & (2) & (3) & (4) & (5) & (6) & (7) & (8) & (9)& (10) \\
  Name     & RA$_\text{J2000}$& DEC$_\text{J2000}$& RA$_{opt}$& DEC$_{opt}$& g'& r'& g'-r'& z$_\text{spec}$& S$_M$\\               
           &    hms         &   $^\circ$ ' "           &    hms &     $^\circ$ ' " &           &      &   &   & mJy&\\
\hline
\\
  J1131--1959 & 11:31:09.25 & -19:59:19.0 & 11:31:09.25 & -19:59:19.9 & 23.34 & 22.26 & 1.07 & -& $4.20\pm0.08$\\
  \bf{J1131--2000} & \bf{11:31:13.74} & \bf{-20:00:20.5} & \bf{11:31:13.73} & \bf{-20:00:21.5} & \bf{21.35} & \bf{19.95} & \bf{1.40} & \bf{0.303} & $1.95\pm0.11$\\
  J1131--1947 & 11:31:22.48 & -19:47:31.2 & 11:31:22.47 & -19:47:32.8 & 25.55 & 23.21 & 2.33 & -& $0.47\pm0.09$\\
  J1131--2001 & 11:31:24.77 & -20:01:09.5 & 11:31:24.75 & -20:01:10.6 & 25.55 & 24.15 & 1.39 & -& $10.30\pm0.10$\\
  J1131--1953 & 11:31:26.89 & -19:53:26.1 & 11:31:26.90 & -19:53:27.3 & 21.98 & 21.95 & 0.02 & 3.020& $3.91\pm0.08$\\
  J1131--1951 & 11:31:27.51 & -19:51:46.1 & 11:31:27.52 & -19:51:47.8 & 25.55 & 23.79 & 1.75 & -& $1.36\pm0.09$\\
  J1131--1946 & 11:31:32.53 & -19:46:59.4 & 11:31:32.52 & -19:47:00.0 & 22.01 & 20.78 & 1.22 & -& $0.63\pm0.07$\\
  J1131--1952 & 11:31:43.92 & -19:52:54.3 & 11:31:43.84 & -19:52:45.8 & 20.24 & 18.89 & 1.34 & 0.257& $16.20\pm0.10$\\
  \bf{J1131--1954} & \bf{11:31:46.79} & \bf{-19:54:52.5} & \bf{11:31:47.13} & \bf{-19:54:52.7} & \bf{21.05} & \bf{19.63} & \bf{1.41} & \bf{0.302} & $0.25\pm0.05$\\
  J1131--1950 & 11:31:48.42 & -19:50:43.8 & 11:31:48.41 & -19:50:45.2 & 23.33 & 22.64 & 0.69 & -& $0.70\pm0.08$\\
  \bf{J1131--1949} & \bf{11:31:48.62} & \bf{-19:49:01.7} & \bf{11:31:48.56} & \bf{-19:49:02.2} & \bf{21.23} & \bf{20.21} & \bf{1.02} & \bf{0.302} & $0.25\pm0.05$\\
  \bf{J1131--1958} & \bf{11:31:49.50} & \bf{-19:58:07.9} & \bf{11:31:49.53} & \bf{-19:58:07.6} & \bf{20.87} & \bf{19.74} & \bf{1.13} & \bf{0.295} & $0.39\pm0.05$\\
  \bf{J1131--1953b} & \bf{11:31:54.30} & \bf{-19:53:53.5} & \bf{11:31:54.27} & \bf{-19:53:50.8} & \bf{20.86} & \bf{19.41} & \bf{1.45} & \bf{0.305} & $37.80\pm0.40$\\
  \bf{J1131--1955} & \bf{11:31:54.34} & \bf{-19:55:39.0} & \bf{11:31:54.18} & \bf{-19:55:39.8} & \bf{20.09} & \bf{18.62} & \bf{1.46} & \bf{0.307} & $12.10\pm0.16$\\
  \bf{J1131--1952b} & \bf{11:31:54.85} & \bf{-19:52:07.7} & \bf{11:31:54.95} & \bf{-19:52:10.2} & \bf{20.70} & \bf{19.24} & \bf{1.45} & \bf{0.303} & $2.30\pm0.13$\\
  J1131--1959 & 11:31:55.60 & -19:59:33.2 & 11:31:55.46 & -19:59:34.2 & 25.53 & 23.12 & 2.43 & 0.740& $0.40\pm0.17$\\
  J1131--2002 & 11:31:57.47 & -20:02:30.5 & 11:31:57.40 & -20:02:32.1 & 21.93 & 20.78 & 1.14 & -& $0.35\pm0.15$\\
  J1131--1946 & 11:31:58.17 & -19:46:56.5 & 11:31:58.19 & -19:46:58.2 & 22.09 & 20.42 & 1.66 & 0.227& $0.29\pm0.10$\\
  \bf{J1132--1954} & \bf{11:32:02.77} & \bf{-19:54:09.0} & \bf{11:32:02.70} & \bf{-19:54:13.5} & \bf{21.23} & \bf{19.78} & \bf{1.44} & \bf{0.306} & $1.02\pm0.10$\\
  \bf{J1132--1952} & \bf{11:32:04.16} & \bf{-19:52:11.1} & \bf{11:32:04.41} & \bf{-19:52:11.1} & \bf{21.14} & \bf{20.39} & \bf{0.74} & \bf{0.302} & $0.25\pm0.05$\\
  J1132--1956 & 11:32:13.96 & -19:56:09.2 & 11:32:14.02 & -19:56:10.9 & 25.55 & 23.26 & 2.29 & -& $0.69\pm0.09$\\
  J1132--1951 & 11:32:21.71 & -19:51:41.3 & 11:32:21.71 & -19:51:42.9 & 25.55 & 23.56 & 1.98 & -& $0.86\pm0.10$\\
  J1132--1949 & 11:32:24.85 & -19:49:01.1 & 11:32:24.84 & -19:49:02.3 & 24.04 & 22.19 & 1.84 & -& $0.50\pm0.19$\\
  J1132--1948 & 11:32:30.09 & -19:48:47.0 & 11:32:30.08 & -19:48:48.4 & 23.68 & 21.96 & 1.72 & -& $0.72\pm0.08$\\
  J1132--1948 & 11:32:31.26 & -19:48:40.0 & 11:32:31.30 & -19:48:41.8 & 23.62 & 21.87 & 1.74 & -& $2.17\pm0.22$\\
  J1132--1952 & 11:32:34.77 & -19:52:43.5 & 11:32:34.73 & -19:52:45.0 & 18.27 & 18.24 & 0.03 & 0.389& $0.25\pm0.11$\\
  \\
\hline
\end{tabular}
\end{center}
\end{footnotesize}
\end{table}

\bsp	
\label{lastpage}
\end{document}